\documentclass[preprint]{aastex}
\pdfoutput=1 
\usepackage[usenames,dvipsnames]{color}
\usepackage{natbib,graphicx,latexsym,lscape,pdflscape,url}
\usepackage[breaklinks,colorlinks,urlcolor=blue,citecolor=black,linkcolor=blue]{hyperref} 

\newcommand{\Kepler}{{\sl Kepler}}

\newcommand{\Hipparcos}{{\sl Hipparcos}}

\newcommand{\Msun}{\mbox{$M_{\sun}$}}
\newcommand{\Mearth}{\mbox{$M_{\oplus}$}}
\newcommand{\Rearth}{\mbox{$R_{\oplus}$}}

\newcommand{\Mjup}{\mbox{$M_{\rm Jup}$}}

\newcommand{\degree}{\mbox{$^{\circ}$}}
\newcommand{\perpix}{\mbox{pixel$^{-1}$}}

\newcommand{\ms}{\hbox{m\,s$^{-1}$}}
\newcommand{\msyr}{\hbox{m\,s$^{-1}$\,yr$^{-1}$}}
\newcommand{\kms}{\mbox{km\,s$^{-1}$}}
\newcommand{\degyr}{\hbox{deg\,yr$^{-1}$}}
\newcommand{\masyr}{\hbox{mas\,yr$^{-1}$}}

\newcommand{\Kc}{\mbox{$K_{\rm cont}$}}
\newcommand{\Kp}{\mbox{$K^{\prime}$}}

\newcommand{\Mtot}{\mbox{$M_{\rm tot}$}}

\newcommand{\Mstar}{\mbox{$M_{\star}$}}

\newcommand{\leff}{\mbox{$\lambda_{\rm eff}$}}

\newcommand{\obj}{Kepler-444}

\shorttitle{Orbital Architectures: \obj}
\shortauthors{Dupuy et al.}

\begin{document}

\title{Orbital Architectures of Planet-Hosting Binaries: \\
I. Forming Five Small Planets in the Truncated Disk of Kepler-444A\altaffilmark{*}}

\author{Trent J.\ Dupuy,\altaffilmark{1}
        Kaitlin M.\ Kratter,\altaffilmark{2}
        Adam L.\ Kraus,\altaffilmark{1}
        Howard Isaacson,\altaffilmark{3}
        Andrew W.\ Mann,\altaffilmark{1} 
        Michael J.\ Ireland,\altaffilmark{4}
        Andrew W.\ Howard,\altaffilmark{5} and
        Daniel Huber\altaffilmark{6}}

      \altaffiltext{*}{Data presented herein were obtained at the
        W.M.\ Keck Observatory, which is operated as a scientific
        partnership among the California Institute of Technology, the
        University of California, and the National Aeronautics and
        Space Administration. The Observatory was made possible by the
        generous financial support of the W.M.\ Keck Foundation.}

      \altaffiltext{1}{The University of Texas at Austin, Department
        of Astronomy, 2515 Speedway C1400, Austin, TX 78712, USA}

      \altaffiltext{2}{Department of Astronomy, University of Arizona,
        933 N Cherry Ave, Tucson, AZ, 85721}

      \altaffiltext{3}{Astronomy Department, University of California,
        Berkeley, CA 94720, USA}

      \altaffiltext{4}{Research School of Astronomy \& Astrophysics,
        Australian National University, Canberra ACT 2611, Australia}

      \altaffiltext{5}{Institute for Astronomy, University of Hawaii,
        2680 Woodlawn Drive, Honolulu HI 96822}

      \altaffiltext{6}{Sydney Institute for Astronomy, School of
        Physics, University of Sydney, NSW 2006, Australia}

\begin{abstract}

  We present the first results from our Keck program investigating the
  orbital architectures of planet-hosting multiple star systems. \obj\
  is a metal-poor triple star system that hosts five sub-Earth-sized
  planets orbiting the primary star (\obj{A}), as well as a spatially
  unresolved pair of M dwarfs (\obj{BC}) at a projected distance of
  1$\farcs$8 (66\,AU).  We combine our Keck/NIRC2 adaptive optics
  astrometry with multi-epoch Keck/HIRES RVs of all three stars to
  determine a precise orbit for the BC pair around A, given their
  empirically constrained masses.  We measure minimal astrometric
  motion ($1.0\pm0.6$\,\masyr, or $0.17\pm0.10$\,\kms), but our RVs
  reveal significant orbital velocity ($1.7\pm0.2$\,\kms) and
  acceleration ($7.8\pm0.5$\,\msyr). We determine a highly eccentric
  stellar orbit ($e=0.864\pm0.023$) that brings the tight M~dwarf pair
  within $5.0^{+0.9}_{-1.0}$\,AU of the planetary system. We validate
  that the system is dynamically stable in its present configuration
  via n-body simulations.  We find that the A--BC orbit and planetary
  orbits are likely aligned (98\%) given that they both have edge-on
  orbits and misalignment induces precession of the planets out of
  transit.  We conclude that the stars were likely on their current
  orbits during the epoch of planet formation, truncating the
  protoplanetary disk at $\approx$2\,AU.  This truncated disk would
  have been severely depleted of solid material from which to form the
  total $\approx$1.5\,\Mearth\ of planets.  We thereby strongly
  constrain the efficiency of the conversion of dust into planets and
  suggest that the \obj\ system is consistent with models that explain
  the formation of more typical close-in \Kepler\ planets in normal,
  not truncated, disks.

\end{abstract}

\keywords{astrometry --- binaries: close --- planetary systems ---
  stars: individual (\obj)}


\section{Introduction}

The demographics of planets around single stars have been studied
extensively for two decades, but statistical studies of planets in
binary systems have been hampered by observational selection effects.
The \Kepler\ prime mission \citep{2010Sci...327..977B} has produced
the first large sample of planet detections that are minimally biased
with respect to the multiplicity of the stellar hosts.  While \Kepler\
data has been used to identify a number of circumbinary planets
\citep[e.g.,][]{2011Sci...333.1602D, 2012Natur.481..475W},
high-angular resolution surveys of \Kepler\ planet candidate hosts
have been needed to discover so called ``s-type'' planets that orbit
only one star in a multiple star system.  Such work has aided in
planet validation \citep[e.g.,][]{2012AJ....144...42A,
  2013AJ....146....9A, 2014A&A...566A.103L, 2015AJ....149...55E}, and
studies of the binary frequency in the \Kepler\ planet sample have
also provided the first constraints on how planet occurrence is
affected by the presence of stellar companions
(\citealp{2014ApJ...791..111W, 2015arXiv151001964W}; Kraus et al.,
submitted to AJ).  However, the impact of stellar multiplicity on
planet formation and system architecture depends not simply on the
presence of a companion but on orbital parameters such as eccentricity
and mutual inclination.

Theoretical work on the formation of planets in binaries preceded
their discovery by many decades \citep[e.g.,][]{1974Icar...22..436H}.
Companion stars are expected to both excite the random velocities of
planetesimals, inhibiting their growth to large sizes, and truncate
protoplanetary disks due to tidal effects
\citep{Artymowicz:1994}. These effects were long thought to inhibit
planet formation in binary systems where the semimajor axis is
comparable to the disk radius. Indeed, observations of the occurrence
of protoplanetary disks among single stars versus binaries has shown
that disks exist but are much less common among $<$40\,AU binaries
\citep[e.g.,][]{1997ApJ...490..353G, 2001ApJ...556..265W,
  2009ApJ...696L..84C, 2011ApJ...731....8K, 2012ApJ...745...19K}.  In
addition, \citet{2012ApJ...751..115H} showed that when the individual
components of binaries host protoplanetary disks, their masses are
depleted by a factor of $\sim$5 for 30--300\,AU binaries and $\sim$25
for $<$30\,AU binaries. Despite such hostile factors, some planet
hosting binary systems exist, e.g., $\gamma$~Cep that has a 20\,AU
binary semimajor axis and a 2\,\Mjup\ planet at 2\,AU
\citep{2003ApJ...599.1383H}.  Theoretical studies of the $\gamma$~Cep
system have shown that even the truncated disk mass was plausibly
large enough to accommodate giant planet formation
\citep{2008ApJ...683L.191J}.  More general theoretical work
\citep{Rafikov:2015, 2015ApJ...798...71S} demonstrates that planet
formation may occur in binaries if planetesimals are either large, as
expected from streaming instabilities \citep{Youdin:2005}, or if the
birth disk is massive enough to damp random planetesimal velocities.
However, these successful theories do not anticipate the extreme
dynamical environment we describe here.

We present the first results of our campaign to observationally
determine the orbital architectures of the stellar multiples that host
\Kepler\ planets by measuring their orbital motion.  \obj\ (a.k.a.\
BD+41~3306, HIP~94931, KOI-3158) was discovered by
\citet{2014A&A...566A.103L} to have a 1$\farcs$8 companion,
corresponding to a projected separation of 66\,AU given the
\Hipparcos\ parallactic distance of $35.7^{+1.0}_{-1.1}$\,pc
\citep{2007hnrr.book.....V}.  \citet{2015ApJ...799..170C} validated
that the five \Kepler\ candidates are sub-Earth radius planets
(0.40--0.74\,\Rearth) orbiting the primary star \obj{A} and reported
that the companion is in fact a double-lined spectroscopic binary
itself, which we denote \obj{BC}.  \citet{2015ApJ...799..170C} also
used asteroseismology of \obj{A} to derive stellar parameters
including system an age of $11.2\pm1.0$\,Gyr, which they note is
consistent with kinematic and compositional evidence for the star
being a member of the thick disk.  While three other hierarchical
triple systems are known to host planets, the stellar companions have
much larger projected separations from the host stars
\citep[240--330\,AU;][]{2014ApJ...788....2B, 2015arXiv151000015E} and
no others are known to host multi-planet systems.

In this paper we combine astrometric and radial velocity (RV)
observations to show that the orbit of the M~dwarf pair \obj{BC}'s
center of mass around the planet hosting star \obj{A} (hereinafter
called the ``A--BC orbit'') is highly eccentric with a closest
approach of 5\,AU.  We discuss why this orbital configuration is
likely primordial and how it is expected to have strongly impacted the
protoplanetary disk that formed the planetary system around
\obj{A}. We review plausible formation scenarios for this peculiar
system.


\section{Observations \label{sec:obs}}

\subsection{Keck/NIRC2 AO Imaging \label{sec:nirc2}}

\begin{figure}
\centerline{
\includegraphics[width=6.5in,angle=0]{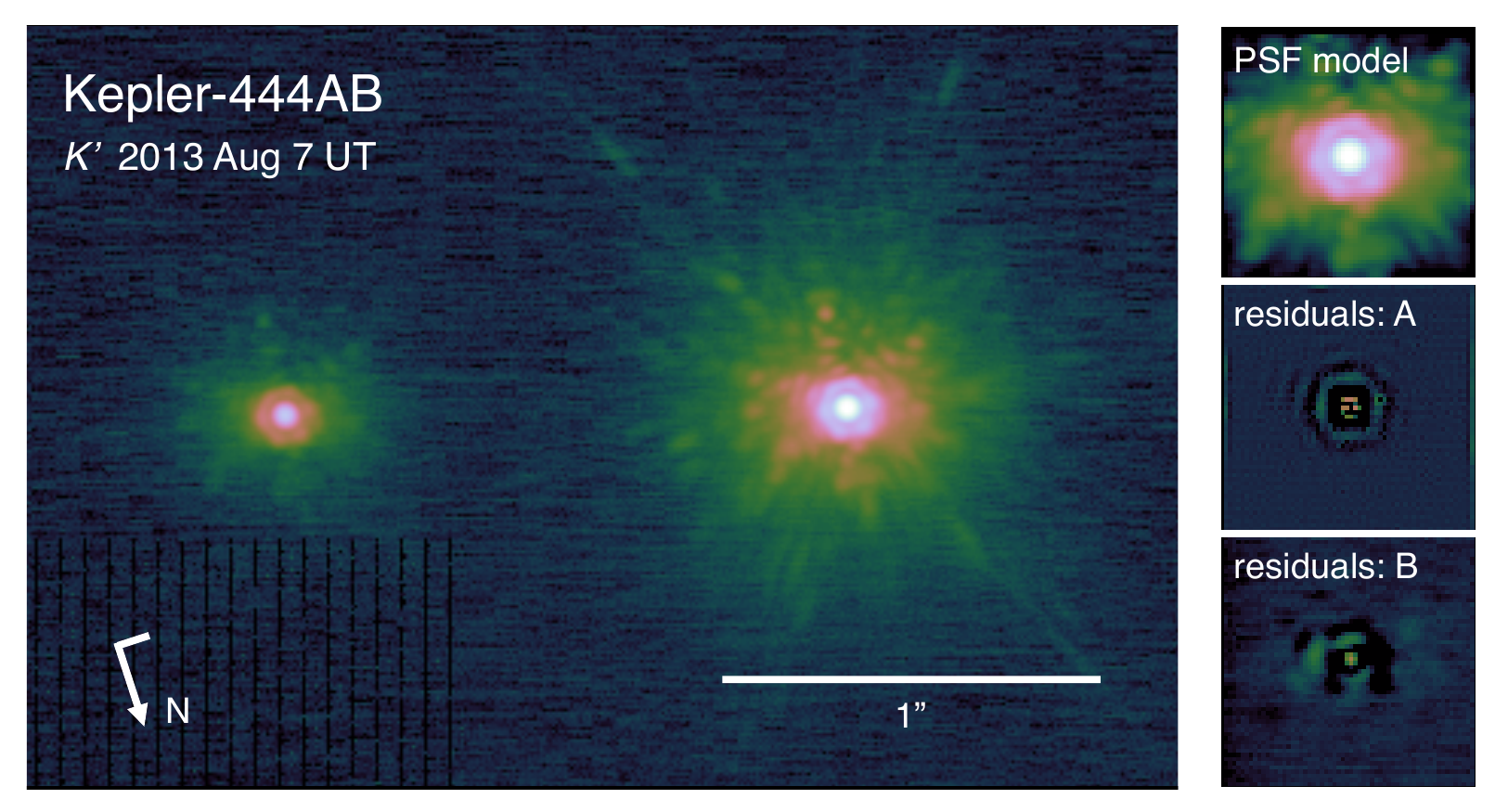}
}
\caption{\normalsize Left: an example of one of our Keck/NIRC2 NGS AO
  images of \obj{AB}. Right: results of our StarFinder PSF-modeling of
  this image showing the best-fit PSF model (top), and residuals when
  subtracting this from the primary (middle) and secondary (bottom)
  components.  \obj{B} is known to be a double-lined spectroscopic
  binary, but we do not resolve it ($<$10\,mas) at any of our four
  observation epochs.  \label{fig:keck}}
\end{figure}

We monitored the \obj\ system with the natural guide star (NGS)
adaptive optics (AO) system at Keck~II \citep{2000PASP..112..315W}
from 2013~Aug~7~UT to 2015~Apr~11~UT.  We obtained data with the
facility imager NIRC2 in both the standard Mauna Kea Observatories
\Kp\ band and the narrow-band filter \Kc\ ($\lambda_{\rm cent} =
2.2705$\,\micron, $\Delta\lambda = 0.0296$\,\micron).  Our images were
reduced in a standard fashion, performing linearity correction, bias
subtraction, flat fielding, and correction of bad pixels and cosmic
rays as described in Kraus et al.\ (submitted to AJ).  We used
StarFinder PSF-fitting \citep{2000A&AS..147..335D} to measure the
precise positions of the two components, as in our previous work on
Keck AO imaging of binaries \citep[e.g.,][]{2009ApJ...692..729D,
  2010ApJ...721.1725D, 2014ApJ...790..133D}.  Examining the residuals
of our fits we found no evidence for any other resolved components in
the system (Figure~\ref{fig:keck}).  Thus we place a limit of
$\approx$10\,mas on the separation of the \obj{BC} pair at all three
epochs, implying a semimajor axis of $\lesssim$0.3\,AU.

To convert $(x,y)$ measurements in individual exposures to positions
on the sky we first corrected for the optical distortion of NIRC2
using the \citet{2010ApJ...725..331Y} calibration, applying their
pixel scale of 9.952\,mas\,\perpix\ and +0\fdg252 correction for the
orientation given in the NIRC2 image headers.  Our NIRC2 data from
2013 and 2014 were obtained in vertical angle mode, where the sky
rotation of the images is constantly changing, so we corrected the
rotator angles reported in those headers to correspond to the midpoint
instead of the start of the exposure.
We then applied corrections for differential aberration and
atmospheric refraction.  The refraction correction requires knowledge
of the air temperature, pressure, and humidity on Mauna Kea during our
observations, for which we used the weather data archived by the
Canada-France-Hawaii
Telescope.\footnote{\url{http://mkwc.ifa.hawaii.edu/archive/wx/cfht/}}
In \Kp\ band, this refraction correction varies slightly between
\obj{A} and \obj{B} because their $K$-band spectra result in a small
difference in the effective wavelength ($\leff$) of our observations.
To compute these wavelengths we convolved the NIRC2 \Kp-band filter
curve\footnote{\url{http://www2.keck.hawaii.edu/inst/nirc2/filters.html}}
with spectra representative of the two components.  We used the IRTF
Spectral
Library\footnote{\url{http://irtfweb.ifa.hawaii.edu/~spex/IRTF_Spectral_Library/}}
spectrum of HD~145675 (K0V) as a template for \obj{A} and the actual
observed spectrum of \obj{B} from IRTF/SpeX (A.\ Mann private
communication, 2015) and found flux-weighted average wavelengths of
$\leff = 2.1089$\,\micron\ and 2.1184\,\micron, respectively.

Table~\ref{tbl:astrom} gives our final separation and position angle
(PA) measurements from NIRC2.  The errors quoted in
Table~\ref{tbl:astrom} are simply the rms of measurements obtained
from multiple individual exposures, but there are other potential
sources of systematic error in this astrometry.  The pixel scale and
orientation have been shown at times to vary between epochs at the
0.002\,mas\,\perpix\ and 0\fdg009 level.\footnote{Examining Table~4 in
  \citet{2010ApJ...725..331Y} shows that given their measurement
  uncertainties epoch-to-epoch variations in scale and orientation are
  indeed significant in a $\chi^2$ sense, where $p(\chi^2) =
  2.8\times10^{-3}$ for the pixel scale and $p(\chi^2) =
  1.0\times10^{-6}$ for the orientation.  In contrast,
  \citet{2010ApJ...725..331Y} performed tests to show that there is no
  evidence of time variations in their distortion solution.}  The
scale uncertainty would correspond to a systematic error of 0.37\,mas
in separation here.  In addition, it is possible that photocenter
shifts due to the orbital motion of the components of \obj{BC} about
their center of mass could affect our (unresolved) NIRC2 astrometry of
this source, although as we derive in Section~\ref{sec:hires} this is
expected to be a relatively small ($<$0.3\,mas) effect.

\begin{figure}
\centerline{\includegraphics[width=3.25in,angle=0]{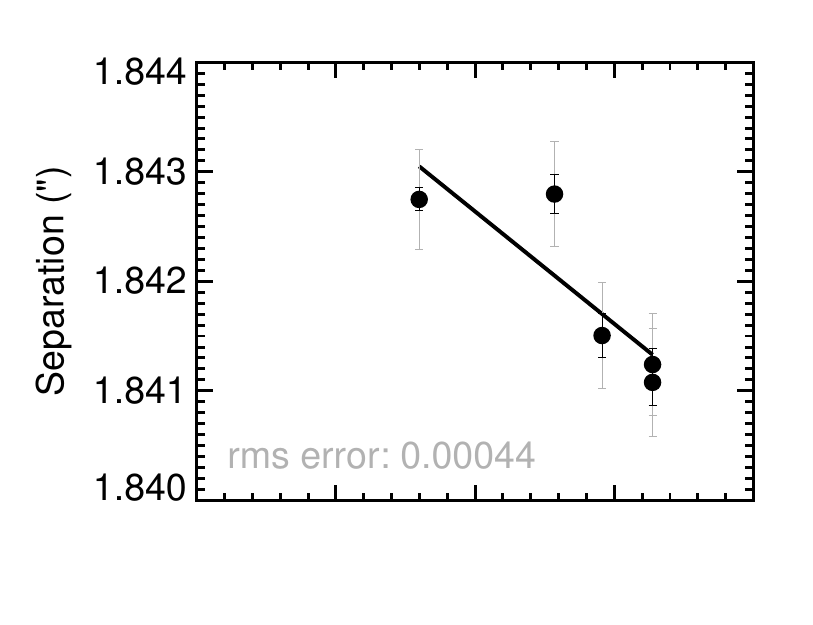}}
\vskip -0.65in
\centerline{\includegraphics[width=3.25in,angle=0]{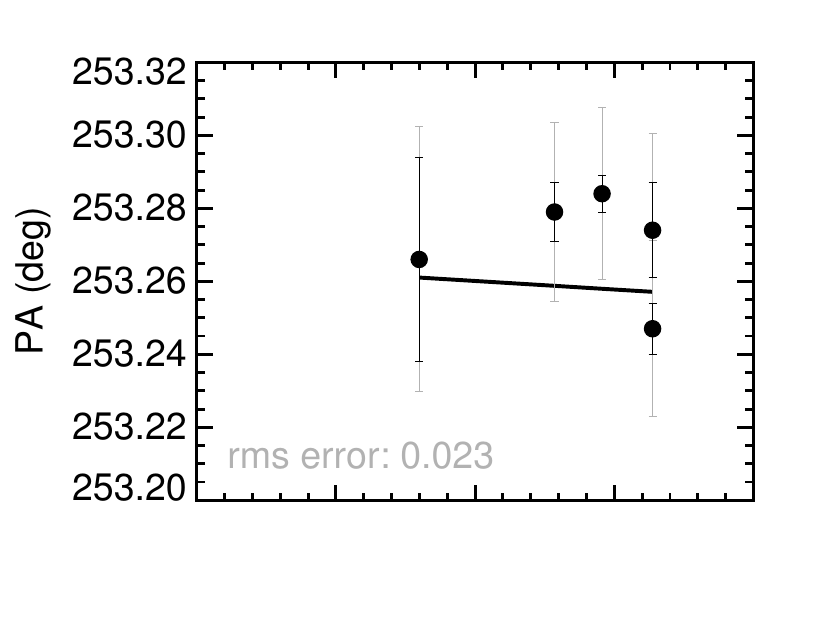}}
\vskip -0.65in
\centerline{\includegraphics[width=3.25in,angle=0]{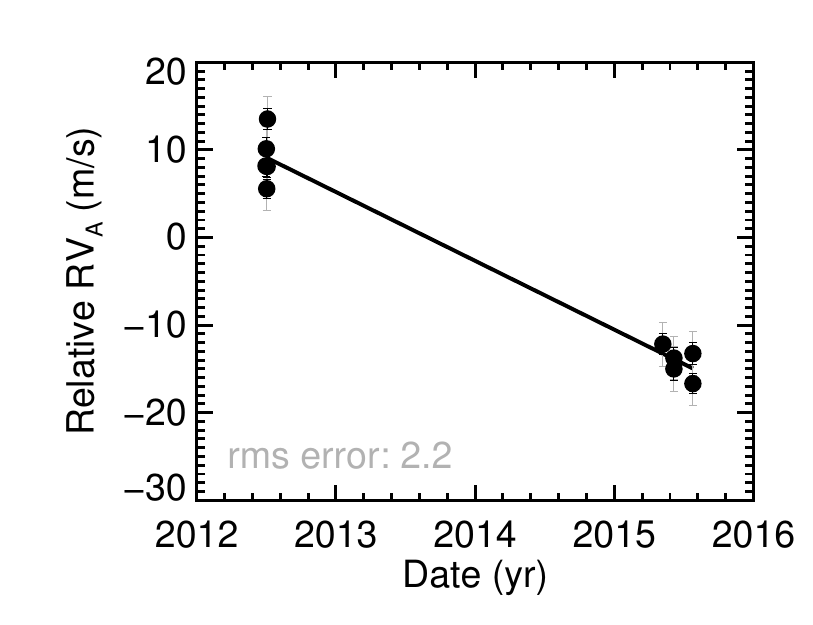}}
\caption{\normalsize Astrometric and RV monitoring data shown
  alongside linear fits to the data as a function of time (see
  Table~\ref{tbl:mcmc} for the coefficients).  In each plot the
  smaller black error bars are the nominal errors (computed from the
  rms of individual dithers for astrometry).  The larger gray error
  bars are the total errors computed from the rms about the fit, which
  should include for example errors in astrometric calibration of
  NIRC2.  The value of this rms error is given in the bottom left of
  each panel.  The astrometry is nearly stationary, indicating very
  little motion in the plane of the sky ($\lesssim$1\,\masyr;
  $\lesssim$0.2\,\kms), while RV monitoring of the planet hosting
  primary star reveals an acceleration in the orthogonal direction due
  to A--BC orbital motion. \label{fig:data}}
\end{figure}

Our relative astrometry for \obj{AB} shows only marginal evidence for
orbital motion.  Thus, to assess our total astrometric uncertainties,
we fit linear relations to both separation and position angle (PA) as
a function of time and adopted the standard deviation (rms) of the
residuals as our final errors.  We find an rms about of the fit of
0.44\,mas in separation and 0\fdg023 in PA, consistent with known
sources of systematic error discussed above.  We find a slight trend
of decreasing separation ($-1.0\pm0.3$\,\masyr) and no evidence for
change in PA ($-0.002\pm0.017$\,\degyr), as shown in
Figure~\ref{fig:data}.  Combining these two linear trends and their
uncertainties, the total astrometric motion is $1.0\pm0.6$\,\masyr.
At a distance of 35.7\,pc, these marginal astrometric motion
detections correspond to $0.17\pm0.05$\,\kms\ in separation and
$0.01\pm0.09$\,\kms\ in PA (total motion of $0.17\pm0.10$\,\kms).
This is much smaller than, for example, the 4.2\,\kms\ relative
velocity of a circular 66\,AU binary with $\Mtot=1.3$\,\Msun.  We also
note that this almost negligible change in the relative astrometry
between \obj{A} and \obj{BC} confirms that this is a physically bound
system, since the primary star's proper motion is 640\,\masyr.

\subsection{Keck/HIRES RVs \label{sec:hires}}

We obtained spectra of \obj{A} with the HIRES spectrometer on the
Keck~I Telescope from 2012~July to 2015~July. The standard setup of
the California Planet Search \citep{2010ApJ...721.1467H} was used in
order to maintain high precision of the radial velocities. During our
first observation of \obj{A}, we identified the unresolved M~dwarf
companion \obj{BC} $\approx$2$\arcsec$ away from the primary. At all
epochs we positioned the decker such that light from the companion did
not contaminate the spectra of the primary. If the seeing conditions
deteriorated to greater than $1\farcs0$ and the two stars could not be
isolated from each other, then the star was not observed.  Each
observation was taken through the gas cell of molecular iodine (I$_2$)
with typical exposure times of 300\,s. An iodine free exposure was
also obtained in order to compute radial velocities with the forward
modeling technique as described in \citet{2006ApJ...646..505B}.  We
report our relative RVs for \obj{A} in Table~\ref{tbl:rva}, and we
found an absolute RV of $-121.4\pm0.1$\,\kms\ for \obj{A}.  We fit the
relative RVs as a function of time and found a linear trend of
$-7.8\pm0.5$\,\msyr, where the uncertainty is derived adopting the rms
of the fit (2.2\,\ms) as the individual measurement error
(Figure~\ref{fig:data}).  This is consistent with, though somewhat
smaller in amplitude, than the linear trend of $-11\pm5$\,\kms\
reported by \citet{2009ApJ...697..544S}.  In the following orbital
analysis we use both our new, precise RV trend along with this value
from the literature.

\begin{figure}
\centerline{\includegraphics[width=6.5in,angle=0]{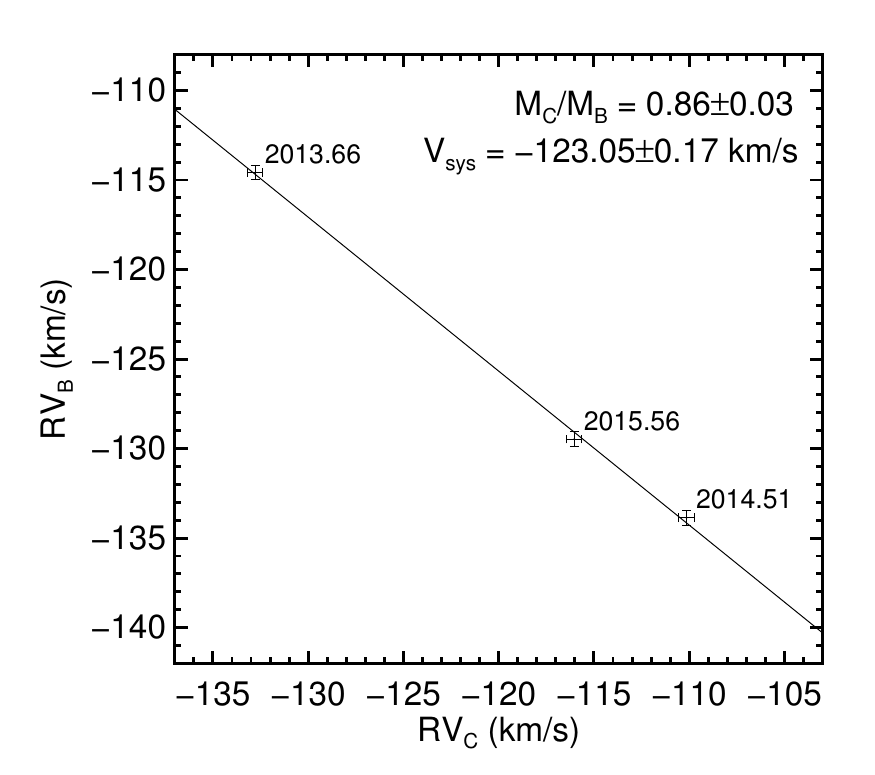}}
\caption{\normalsize Radial velocities for the spectroscopic binary
  companion system \obj{BC}.  Even without a full orbit fit, a linear
  fit to RV$_{\rm C}$ as a function of RV$_{\rm B}$ is sufficient to
  determine the system velocity and mass ratio
  \citep{1941ApJ....93...29W}.  By combining this system velocity
  ($RV_{\rm BC} = -123.05\pm0.17$\,\kms) with the known RV of the
  primary (RV$_{\rm A} = -121.4\pm0.1$\,\kms) we derive a $\Delta$RV
  of $-1.7\pm0.2$\,\kms.  We therefore detect significant orbital
  motion orthogonal to the plane of the sky, even though our
  astrometry shows almost no motion ($\lesssim$1\,\masyr;
  $\lesssim$0.2\,\kms). \label{fig:wilson}}
\end{figure}

At three epochs, we also obtained HIRES spectra of the companion
\obj{BC} from which we obtained RVs by cross correlation with the well
studied M~dwarf Gl~699, resulting in two clear peaks.
To extract the individual RVs of the two components of \obj{BC} we
simultaneously fit a two-component Gaussian to each peak in the
cross-correlation functions.  Table~\ref{tbl:rvbc} reports our radial
velocities for the components of \obj{BC}.  To derive the system
velocity of the \obj{BC} barycenter, we fit a linear relation to
RV$_{\rm B}$ as a function of RV$_{\rm C}$
\citep{1941ApJ....93...29W}, and from the rms of the fit residuals we
determine errors of 0.4\,\kms\ in our deblended RVs for \obj{BC}
(Figure~\ref{fig:wilson}).  We thus find a system velocity of
$-123.05\pm0.17$\,\kms\ and mass ratio of $M_{\rm C}/M_{\rm B} =
0.86\pm0.03$ for \obj{BC}.

We estimated the potential amplitude of the \obj{BC} photocenter orbit
in our NIRC2 imaging given this mass ratio.  In our $K$-band imaging
we estimate the binary would have a flux ratio of $0.3\pm0.1$\,mag
based on the 0.4\,mag ratio of cross-correlation function peaks in the
optical RV data.  The size of the photocenter orbit is defined as the
total semimajor axis scaled down by the factor $f-\beta$, where $f
\equiv M_{\rm C}/(M_{\rm B}+M_{\rm C})=0.462\pm0.009$ and $\beta$ is
the ratio of secondary's flux to the total flux ($\beta =
0.432\pm0.023$).  We therefore expect the photocenter orbit to be
$0.030\pm0.024$ times the size of the the semimajor axis.  Since we
detect no elongation at any epoch in our NIRC2 PSF-fitting, the
semimajor axis is most likely to be $<$10\,mas.  We therefore expect
the photocenter orbit to be $<$0.3\,mas, which is smaller than the
epoch-to-epoch uncertainty in the astrometric calibration.


\section{The Highly Eccentric \obj{A}--BC Orbit}

If the \obj{A}--BC orbit has a semimajor axis close to its projected
separation of 66\,AU, its total orbital velocity would (on average) be
4.2\,\kms, or 25\,\masyr, given its distance of 35.7\,pc and a system
mass of 1.30\,\Msun.  In contrast, we detect minimal astrometric
motion ($-1.0\pm0.3$\,\masyr\ in separation, $-0\fdg002\pm0\fdg017$ in
PA) and a relatively small change in radial velocity ($\Delta{\rm
  RV}_{\rm A-BC} = -1.7\pm0.2$\,\kms).  These measurements alone imply
either that the A--BC orbit is eccentric and near apoastron or that it
simply has a semimajor axis much larger than its projected separation.
The detection of significant acceleration in \obj{A}'s radial velocity
favors the eccentric orbit scenario.  To quantify the \obj{A}--BC
orbital parameters, we performed a Markov chain Monte Carlo (MCMC)
joint analysis of our astrometric and RV data.

We used the Python implementation of the parallel-tempering ensemble
sampler in \texttt{emcee~v2.1.0} with 100 walkers and 30 temperatures.
We found that parallel-tempering sampled our orbital parameter space
more efficiently than the affine-invariant sampler
\citep{2013PASP..125..306F} because a very wide range of orbits are
consistent with our nearly stationary astrometry.  We built up the
initial starting points of our chains by iteratively adding in
observational constraints.  We began with a set of orbital parameters
found by performing a separate Monte Carlo analysis of our astrometry,
where we searched 10$^6$ randomly drawn values for orbital period
($P$), eccentricity ($e$), and time of periastron passage ($T_0$).  As
shown by \citet{2014A&A...563A.126L}, combining astrometric
measurements with this set of three parameters allows best-fit values
for the other visual binary parameters to be determined in a
least-squares sense.  A subset of 10$^3$ of these trials having
$\chi^2-\chi_{\rm min}^2 < 1$ were passed along to \texttt{emcee} as
the starting points for our MCMC.  After running \texttt{emcee} for
10$^5$ steps we added in the $\Delta{\rm RV}_{\rm A-BC}$ constraint,
then after another 10$^5$ steps added in the RV linear trend
measurements.

In our analysis, we fixed the distance at 35.7\,pc but allowed for an
uncertainty in the system mass of $1.30\pm0.06$\,\Msun.  The
asteroseismic analysis of \citet{2015ApJ...799..170C} determined the
mass of \obj{A} ($0.76\pm0.04$\,\Msun), and we use the mass--magnitude
relation from \citet{2000A&A...364..217D} to estimate masses of
$0.29\pm0.03$\,\Msun\ and $0.25\pm0.03$\,\Msun\ for \obj{B} and
\obj{C} based on their absolute magnitudes of $M_K = 6.91$\,mag and
7.21\,mag, respectively.  The mass ratio of the A--BC system was fixed
in our analysis to be $(M_{\rm B}+M_{\rm C})/M_{\rm A} = 0.71$. We
adopted uniform priors in the logarithm of semimajor axis ($\log{a}$),
eccentricity ($e$), argument of periastron ($\omega$), PA of the
ascending node ($\Omega$), and mean longitude at the reference epoch
of 2456511.83~JD ($\lambda_{\rm ref}$).  We assumed randomly
distributed viewing angles by adopting an inclination prior uniform in
$\cos{i}$.
Finally, we imposed a ``discovery prior'' that was computed as the
probability of detecting the binary companion in our
$10\arcsec\times10\arcsec$ NIRC2 images at a random observation time.
This effectively rules out extremely wide orbits ($>$100\arcsec) that
only appear to have a small angular separation due to an improbable
viewing angle.  For a given semimajor axis (in angular units) this
discovery prior is only a function of $e$, $i$, and $\omega$, so we
interpolated the prior from a look-up table with grid steps of 0.03 in
$e$, 3\,deg in $i$, and 6\,deg in $\omega$.  This prior has a
relatively small impact on the results here, as it only affects
orbital solutions with very large semimajor axes that match our
astrometry well but do not match our RV data.

\begin{landscape}
\begin{figure}
\centerline{\includegraphics[width=8.0in,angle=0]{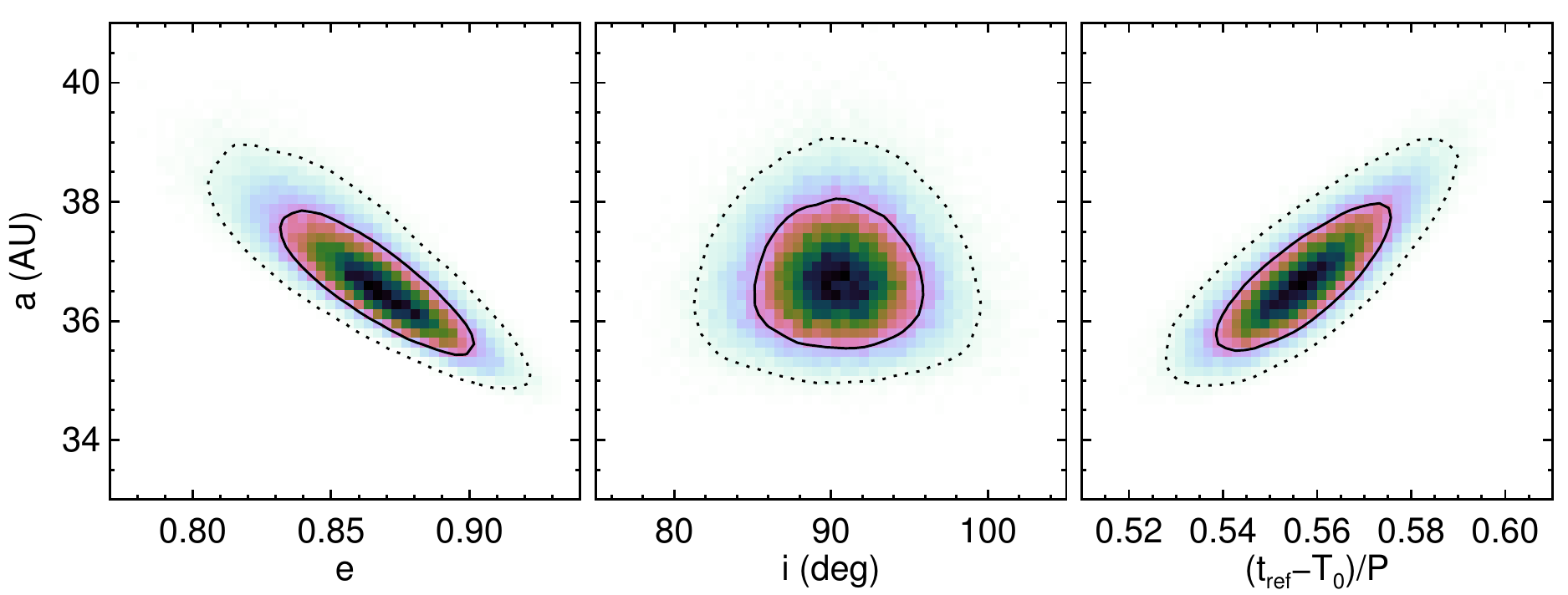}}
\vskip 0.2in
\centerline{\includegraphics[width=8.0in,angle=0]{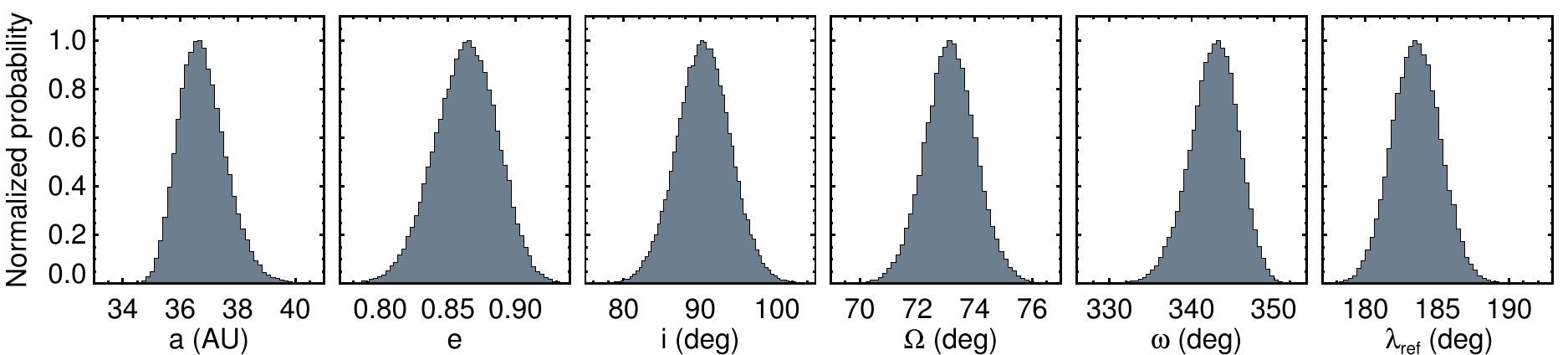}}
\vskip 0.2in
\caption{\normalsize MCMC posterior distributions for the properties
  of the orbit of \obj{BC} around \obj{A}. Top: parameter correlations
  shown as 2-d probability density, with contours indicating 1$\sigma$
  (solid) and 2$\sigma$ (dotted) regions. Bottom: marginalized
  posteriors of all 6 fitted orbital parameters.  \obj{AB} is
  currently at a projected separation of 66\,AU because it is in fact
  a highly eccentric, 37\,AU binary that is near apoastron and that is
  also consistent with being seen edge on (defined as $i=90$\,deg).
  Apoastron is defined as $(t-T_0)/P=0.5$, and we have defined $t_{\rm
    ref}$ as the first epoch of Keck astrometry on 2013~Aug~7~UT.
  Bottom: marginalized posterior distributions for each of the six
  fitted orbital parameters. \label{fig:mcmc}}
\end{figure}
\end{landscape}

Figure~\ref{fig:mcmc} shows the six fitted parameters' posterior
distributions along with the most significant parameter correlations,
and Table~\ref{tbl:mcmc} gives the corresponding credible intervals
for all orbital parameters of interest.  We confirm that the A--BC
orbit is indeed currently near apoastron, where it should spend most
of its time, on a highly eccentric orbit ($e=0.864\pm0.023$).  The
joint constraint from detecting almost no astrometric motion and our
measurement of both velocity and acceleration orthogonal to the plane
of the sky allows a remarkably precise determination of orbital
parameters.  For example, our MCMC gives an inclination of
$i=90.4^{+3.4}_{-3.6}$\,deg.  If we examine the best-fit orbits at
assumed inclinations ranging from $i=80$\,deg to $100$\,deg, those
orbits display PA motion of 0.057\,\degyr\ to $-$0.044\,\degyr.  Our
astrometry rules out such motion at 3.5$\sigma$ and 2.5$\sigma$,
respectively, even though the RVs computed from such orbits are quite
consistent with our measurements.


\section{Discussion}

\subsection{Dynamical Stability \label{sec:stab}}

We have found that the orbit of \obj{BC} brings its center of mass
within $5.0^{+0.9}_{-1.0}$\,AU of \obj{A} and its planetary system.
Both the planetary and stellar orbits are subject to dynamical
instabilities if the orbits are not sufficiently hierarchical.
Extrapolation of empirical fits for stability of s-type planetary
orbits from \citet{1999AJ....117..621H} suggest that the widest
allowed planetary orbit around the primary is $\approx$1.6\,AU, i.e.,
20$\times$ larger than the 0.08\,AU orbit of \obj{f}.  This semi-major
axis is likely an over estimate given that their fits do not consider
hierarchical triples or orbits as eccentric as this.  The existence of
the triple stellar system itself also provides a stability constraint
without considering the planets.  According to
\citet{2006tbp..book.....V}, the triple system is stable as long as
the tight pair \obj{BC} has a semi-major axis less than 1.0\,AU.  This
is is consistent with the fact it was not resolved at any epoch of our
Keck AO imaging ($<0.3$\,AU).

To assess the internal stability of the five-planet system, we
estimated masses for the five planets from the
\citet{2011ApJS..197....8L} mass--radius relation, $M/\Mearth =
(R/\Rearth)^{2.06}$.  \citet{2015ApJ...799..170C} reported radii of
$0.403^{+0.016}_{-0.014}$\,\Rearth,
$0.497^{+0.021}_{-0.017}$\,\Rearth,
$0.530^{+0.022}_{-0.019}$\,\Rearth,
$0.546^{+0.017}_{-0.015}$\,\Rearth, and $0.74\pm0.04$\,\Rearth, in
order from the innermost to outermost planet.  We thereby compute
planet masses of 0.15\,\Mearth, 0.24\,\Mearth, 0.27\,\Mearth,
0.29\,\Mearth, and 0.54\,\Mearth, respectively.  Even though the
implied planet densities are high (7--13\,g\,cm$^{-3}$), we find that
the planets masses are still low enough that they are spaced by 18--29
Hill radii, monotonically increasing outward.  This is roughly twice
the canonical limit at which mean motion resonance overlap drives
multi-planet systems unstable \citep{Chambers:1996}.  Therefore, we
find that the planetary orbits should be quite stable in the absence
of outside influences.  Notably, planet pairs (a,b), (c,d), and (d,e)
all fall near a 5:4 mean motion resonance, with (b,c) close to 4:3.
Given the small masses, and the correspondingly small libration
widths, this may not indicate present day resonant locking.

The high eccentricity of the A--BC stellar orbit pushes the bounds of
the simulations on which published empirical stability fits are based,
so we carried out direct n-body integrations tailored to match the
properties of the \obj\ system.  We used the publicly available
Swifter integrator package \citep{2013ascl.soft03001L} with the 15th
order Gauss-Radau integrator \citep{1985dcto.proc..185E}. All of the
planets were assigned their nominal semimajor axes, fixed star-planet
mass ratios of $10^{-7}$, and non-zero eccentricities and inclinations
less than $10^{-3}$ and 1\degree\ respectively.  We tested a range of
eccentricities for the A--BC orbit, not just our measured value of
$e=0.864\pm0.023$, and for these other values of $e$ we assumed that
the binary is seen at 66\,AU presently because it is at apocenter so
that $a(1+e) = 66$\,AU.  We also tested a range of mutual inclinations
ranging from 0--70\degree. We found that even if the tighter B--C
orbit is as wide as 0.3\,AU, the planets were stable over Myr
timescales for A--BC eccentricities as high as $e=0.95$, i.e.,
corresponding to BC center-of-mass passages within 1.6\,AU of the
outermost planet.  In this case, the planets' eccentricities were
excited to $\approx$$10^{-2}$ by the interaction.  Our n-body
integrations cannot rule out much longer, Gyr-timescale instabilities
due to secular resonance instabilities as this would require
computationally expensive integrations and a broader parameter study
beyond the scope of this work.  More simplified secular models, which
neglect the planetary masses, cannot capture the inclination evolution
of the system \citep[e.g., see ][]{2015arXiv151100944H}. General
relativistic effects and stellar tides, neglected here, may induce
precession of the planets pericenter and tidal locking, but should not
decrease orbital stability.  We therefore conclude that our derived
A--BC orbit is not ruled out by dynamical instabilities in the system.

\subsection{Coplanarity of the Stellar and Planetary Orbits \label{ref:align}}

A direct result from our analysis of the \obj{A--BC} orbit is a
measurement of the inclination, which we find to be consistent with
edge-on ($i=90.4^{+3.4}_{-3.6}$\,deg).  The inclinations measured by
\citet{2015ApJ...799..170C} for the five planets are consistent with
being internally coplanar within their errors, with the two most
precise values being $87.96^{+0.36}_{-0.31}$\,deg for \obj{f} and
$89.1\pm0.5$\,deg for \obj{e}.\footnote{Inclinations in the interval
  $i=0$--90\,deg correspond to counter-clockwise orbits, while the
  interval $i=90$--180\,deg corresponds to clockwise orbits.
  Transiting planets lack the information to distinguish between these
  two cases, unlike astrometric orbits, so transiting planet
  inclinations are reported in the interval $i=0$--90\,deg.  In other
  words, the measurement for \obj{f} could equivalently be interpreted
  as either $i=89.1\pm0.5$\,deg or $i=90.9\pm0.5$\,deg.} The planet
orbits are then also consistent with having the same sky projected
inclinations as the A--BC stellar orbit at $<$1$\sigma$.  However, the
planets could still have some mutual inclination with respect to the
A--BC orbit.  Directly constraining such mutual inclination requires
knowledge of the PA of the transiting planets' orbits (i.e., their
$\Omega$), but this is observationally inaccessible.  We therefore
assess the likelihood of coplanarity from probabilistic arguments.
The probability of having a sky projected inclination $i$ by chance
alone is $p(i)=\sin(i)/2$.  Integrating this function we find a
probability of randomly having an inclination within 4\degree\ of
edge-on is 7\%.  This is the probability of the stellar orbit being
observed with $i=86$--94\degree\ if its mutual inclination with
respect to the planetary orbits were in fact randomly oriented.

Our numerical integrations described in Section~\ref{sec:stab} offer a
second probabilistic constraint on orbital alignments.  When the A--BC
orbit is not coplanar with the planetary orbits, precession is induced
in the planets' orbits. 
We find that the planets precess roughly as a rigid disk and cycle
through states in which none or all of them appear to be transiting
along a single sight line.  We consider an initial configuration where
the planet--BC inclination is entirely in the relative PA, with all
planets transiting. This is a conservative initial condition in the
sense that it will provide the most favorable configuration for
transits.  Because the planets precess like a rigid disk, the
outermost planet is least likely to transit for a given inclination,
and when the outer planet transits the interior planets do as well.
The oscillation timescale of the planetary orbits is order a few
$10^5$--$10^6$\,yr, depending on the inclination.  When the planet--BC
orbit has an mutual inclination of 5\degree, all planets transit
roughly 35\% of the time.  The fraction of time in transit drops to
25\% for a mutual inclination of 10\degree.  Although we do not carry
out an exhaustive parameter study, some larger misalignments can
provide slightly higher transit probabilities because the planetary
orbits can become retrograde, as predicted by
\citet{2014ApJ...785..116L}.  Given our 2$\sigma$ inclination
uncertainty of 7\degree\ and that the five-planet system is
transiting, we determine that the likelihood of misaligned stellar and
planet orbits is $\approx$30\% from this second probabilistic
constraint.

Combining these constraints, that we observe the planetary orbits to
be edge on, the A--BC orbit to be edge on, and the five planets to be
transiting today, we find a probability of only 2\% that the \obj\
A--BC orbit is misaligned with respect to the planets.  Therefore, we
conclude that it is highly likely that the stellar A--BC and planetary
orbits are coplanar within the range of our measurement uncertainties,
providing an important constraint on the origin of the \obj\ system.

Finally, we note that the lack of eclipses between \obj{B} and C in
\Kepler\ light curves does not rule out the \obj{B}--C orbit being
coplanar with the A--BC orbit and planet orbits within the
observational uncertainties.  The lack of eclipses only constrains
their inclination to be $\Delta{i} = 0.5\degree (\frac{R_{\rm
    B}+R_{\rm C}}{0.6\,R_{\odot}}) (\frac{a}{0.3\,{\rm AU}})^{-1}$
away from edge-on ($89.5^{\circ} > i > 90.5^{\circ}$).  Our numerical
integrations indicate that moderate mutual inclinations between the B--C
orbit and the planet orbits would induce only small amplitude
precession ($\lesssim$1\degree) that would not cause the planets to
ever go out of a transiting configuration.

\subsection{Formation of the Triple Star System \label{sec:starform}}

We begin our discussion of plausible formation scenarios for the \obj\
system with the origin of the hierarchical triple stellar system.  One
of the most important questions is whether the triple system is
primordial, by which we mean that it existed in its current orbital
architecture at the epoch of planet formation.  The three components
may have initially formed in a less hierarchical, less eccentric
configuration than observed today, through either core fragmentation
or disk fragmentation.  Observational evidence exists for both
mechanisms (e.g., \citealp{2015Natur.518..213P}; Tobin et al.,
submitted).  However, evolution into the system's current state would
have been rapid.  A violent dynamical interaction could generate an
eccentric triple due to an unstable orbital configuration at birth,
typically occurring within $10^4$ orbits ($\lesssim$$10^6$\,yr;
\citealt{2008IAUS..246..209V}).  In such cases, any interactions of
the triple system with the initial gas cloud or protostellar disks
would be concurrent with planet formation.

To arrive at the current hierarchical stellar arrangement at later
times would most likely require the introduction of a fourth stellar
body.  Even in a relatively populous cluster (10$^3$ stars),
simulations from \citet{2006ApJ...641..504A} indicate that encounters
within 100\,AU after 5\,Myr are rare, occurring at a rate of
$\approx$$3\times10^{-4}$\,star$^{-1}$\,Myr$^{-1}$.  If such an
encounter occurred, the interaction needed to create the tight M~dwarf
pair would likely have required a violent, close passage destroying
the planetary system.  Moreover, the resulting alignment of the
\obj{A}--BC orbit would be expected to be random with respect to the
planetary system around \obj{A}, but we find they are coplanar
(Section~\ref{ref:align}).
Secular instabilities among the current three stars, such as the
Kozai-Lidov mechanism, could cause orbital evolution over long
timescales \citep{1962AJ.....67..591K,1962P&SS....9..719L}. However,
dynamical stability requires that the M~dwarf pair could not have been
significantly wider in the past than it is today.  Finally, even if
there is or was an unknown fourth, wider companion, Kozai oscillations
of the A--BC orbit are disfavored because the precession of the
longitude of the pericenter driven by the A--BC system is faster than
that driven by Kozai oscillations.  We therefore conclude that it is
highly improbable that the formation of the triple stellar system
occurred at late times, so the A--BC orbit is very likely primordial.

\subsection{Formation of the Planetary System \label{sec:planetform}}

Given that the orbit of triple system was likely in place at or before
the epoch of planet formation, we are then confronted with the fact
that the protoplanetary disk would have been truncated by the M~dwarf
pair \obj{BC} on their $a=36.7^{+0.7}_{-0.9}$\,AU, $e=0.864\pm0.023$
orbit about the host star \obj{A}.  Extrapolating from the work of
\citet{Artymowicz:1994}, we expect the disk to be truncated to
1--2\,AU due to the close pericenter passage of 5\,AU.  Correctly
modeling this truncation would require tailored modeling accounting
for the specific triple system architecture here, so instead we
conservatively adopt a truncation radius of 2\,AU for the disk of
\obj{A}.  The supply of solids available for planet formation would be
severely limited in such a disk.  While continued feeding from an
outer circum-triple ring is possible, as seen in systems such as
GG~Tau \citep{2012ApJ...754...72B} or UY~Aur
\citep{2014ApJ...792...56S}, the eccentricity of the A--BC orbit makes
this challenging. Not only would one expect a massive ring to damp the
eccentricity of the orbit, but also the bulk of the material would
accrete onto the BC components, rather than A
\citep{2015MNRAS.452.3085Y}.

\begin{figure}
  \centerline{\includegraphics[width=4.0in,angle=0]{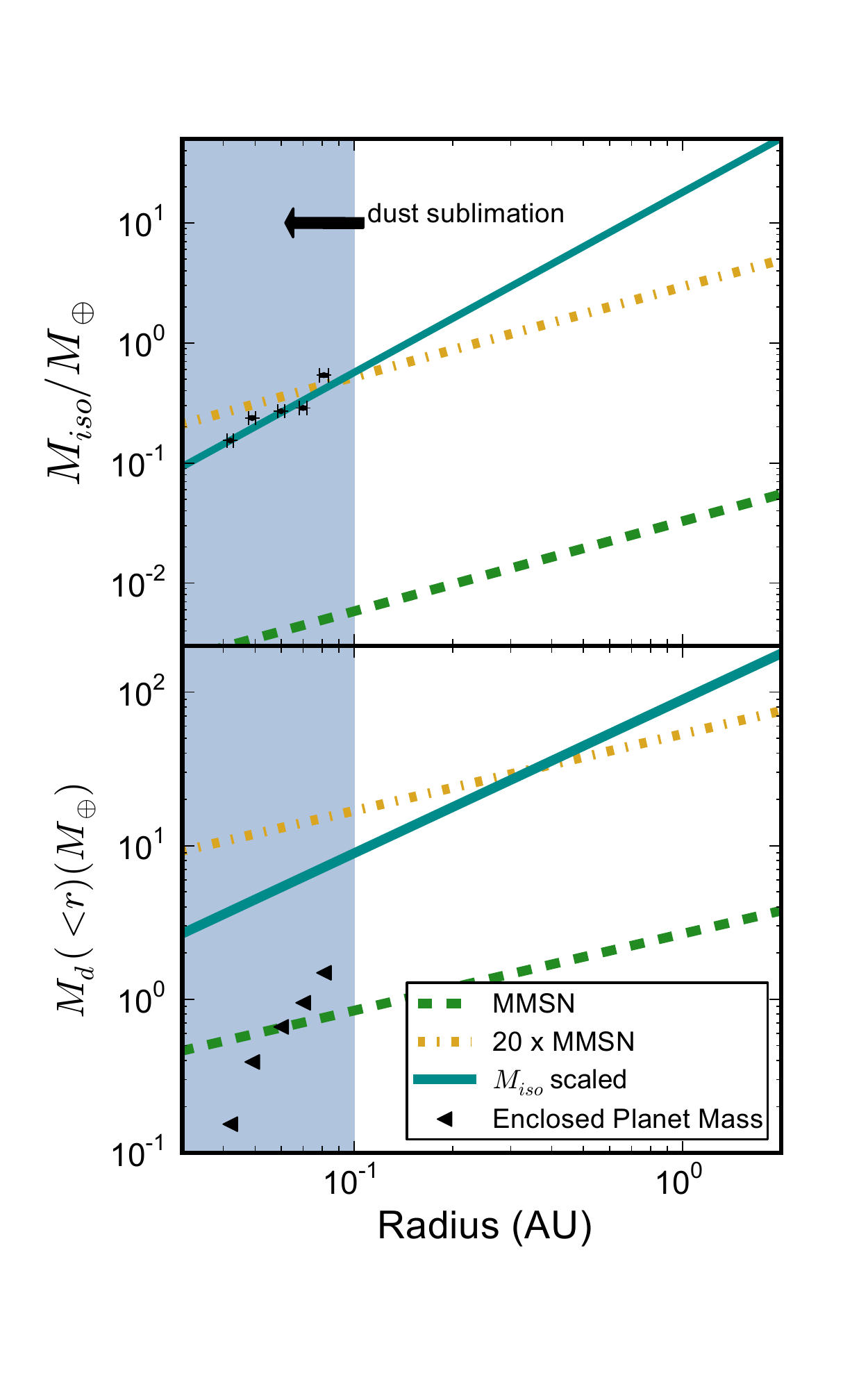}}
  \vskip -0.5in
  \caption{\normalsize Top: comparison of the estimated planet
    masses to the isolation mass of solids ($M_{\rm iso}$) as a
    function of disk radius for three protoplanetary disk models.  The
    green dashed line is a MMSN disk with $\Sigma_{\rm gas} = 1700
    \rm{g/cm}^{-2}$ at 1\,AU, a surface density profile $\Sigma
    \propto r^{-3/2}$, and a dust-to-gas ratio of 1:300. The gold
    dash-dotted line is the same MMSN scaled up by a factor of 20.
    The solid blue line shows that a shallower surface density
    profile, $\Sigma \propto r^{-1}$, produces isolation masses that
    increase more similarly to the planet masses.  The $x$-axis error
    bars on planet masses indicate the size of the feeding zone of
    each planet, demonstrating that though packed, they do not
    overlap. Bottom: the total enclosed mass in these three disk
    models as a function of disk radius.  Black triangles show the
    total enclosed planet mass.  Both the $M_{\rm iso}$-scaled disk
    and the 20$\times$ MMSN models are consistent with disk masses
    that would lead to gravitational instability at
    50--100\,AU. \label{fig:miso}}
\end{figure}

Adopting a minimum mass solar nebula (MMSN) gas surface density of
$\Sigma = 1700\,\rm{g/cm}^{-2} (\frac{r}{\rm 1\,AU})^{3/2}$
\citep{1977Ap&SS..51..153W, 1981PThPS..70...35H}, and a 1:100
dust-to-gas mass ratio, the total mass of solids in \obj{A}'s disk
would have been $\approx$12\,\Mearth.  Given the fact that \obj{A} is
a metal-poor star ([Fe/H]$ = -0.55$\,dex), we might expect the
dust-to-gas ratio to be $\approx$3$\times$ smaller, implying a total
mass budget of $\approx$4\,\Mearth.  Either case would require
remarkably efficient conversion of dust to planets to create the
five-planet system that has an estimated total mass of 1.5\,\Mearth\
(Section~\ref{sec:stab}).
In Figure~\ref{fig:miso} we compare the estimated planet masses to the
predicted isolation mass of solids as a function of radius,
\begin{equation}
  M_{\rm iso } = \frac{(4\pi f_h \Sigma)^{3/2}r^3}{(3\Mstar)^{1/2}},
\end{equation}
where $f_h \approx 3.5$ is a geometric factor \citep{Lissauer:1987}.
The isolation masses for a MMSN disk are far too low to support in
situ formation, and also dust grains should have sublimated at the
planets' current locations.
This is a familiar problem for Kepler systems, particularly for
close-in super-Earths and Neptunes
\citep[e.g.,][]{2012ApJ...751..158H, 2013MNRAS.431.3444C}. For those
more massive planets, local disk masses must be unphysically high
(locally gravitationally unstable) to account for in-situ formation,
implying either drifting in of solids or migration of fully formed
planets \citep{2014ApJ...795L..15S}.  Even with drifting of solids,
disk mass measurements suggest that dust-to-planet conversion
efficiencies may be quite high, depending on the amount of grain
growth that previously occurred during the Class~I phase
\citep{2014MNRAS.445.3315N}.

Current theories posit that close-in rocky planets, even super-Earths,
may have formed near their current orbital locations by delivery of
solids from much larger disk radii \citep{2014ApJ...780...53C,
  2014ApJ...797...95L}. A pressure maximum located at the boundary
between the MRI-active zone and the disk deadzone could provide a
plausible trap for the collection of solids. This deadzone boundary is
thought to occur at $\approx$0.1\,AU and is expected to have weak to
no dependence on metallicity \citep{Martin:2012, 2013ApJ...764...65M}.
At the pressure maximum, pebbles drifting in from large radii collect
and can either coalesce through a ring instability in the solids,
streaming instabilities, or coagulation. Sequential epochs of pebble
gathering could thus produce multiple low mass planets in this
so-called ``inside-out'' planet formation model.

The five \obj\ planets, among the smallest discovered by \Kepler,
appear to be qualitatively consistent with in-situ, drift-aided planet
formation occurring in a truncated disk. The mass reservoir was
severely depleted, which would have resulted in less material being
delivered to the deadzone edge. The material would also likely have
been depleted in volatiles, since the disk would have been truncated
within or very near to the ice-line. Therefore, the planets formed
here might have been expected to be smaller, and denser than those
formed via ``inside-out'' planet formation in a normal disk around a
single star.  The monotonic size ordering is consistent with this
scenario and should not be due to distance dependent mass loss via
oblation \citep{Perez-Becker:2013}, as even the innermost planet has
too low an equilibrium temperature ($<$1500\,K) for this to be
applicable.  The innermost planet orbits roughly a factor of two
closer than the expected formation site at the deadzone boundary.
Type~I migration might have caused inward drift from that formation
site toward the disk's inner edge, as the estimated planet masses
(0.15--0.54\,\Mearth) are well below the gap-opening mass.

Despite these qualitative successes, there is still a quantitative
mass budget problem for a truncated MMSN disk.  
However, unlike most other Kepler systems, the \obj\ planets are small
enough that a disk with only 20$\times$ the surface density of the
MMSN at 1\,AU would have sufficiently high isolation masses to form
the planets locally (albeit at $\sim$100\% efficiency) without relying
on any material from the outer disk. More conservatively, this more
massive disk would contain 80--240\,\Mearth\ of solid material within
2\,AU, depending on the dust-to-gas ratio.
This total supply of solid material would then require only
$\lesssim$2\% efficiency in planet formation, assuming it could be
delivered to the inner disk as described above. 

If the solution to the overall mass budget problem is an unusually
massive disk, this might hint at a coherent formation model for both
the hierarchical triple and the planets.  While a disk 20 times as
massive as the MMSN would be locally stable to gravitational
instabilities, when extrapolated out to typical disk radii of
$\sim$100\,AU \citep{Andrews:2013}, the total disk mass would approach
that of the primary star. At large radii (70--100\,AU) temperatures
are low (40--50\,K), and thus a massive disk would likely became
gravitationally unstable and susceptible to fragmentation
\citep{1989ApJ...347..959A, 2010ApJ...710.1375K}.
Numerical simulations show that disks with such high disk-to-star mass
ratios usually fragment into 1--2 objects that grow to high mass
ratios relative to the host star by successfully competing for
accreting material from the disk and envelope
\citep{Stamatellos:2009a, 2007ApJ...656..959K, 2010ApJ...708.1585K}.
The rapid growth of fragments formed in the outer disk naturally leads
to rapid dynamical evolution with varied outcomes, including ejection,
merging, and inward or outward migration. One possible outcome of
these interactions is an eccentric close binary pair orbiting the
primary, resembling the \obj{ABC} system \citep{Stamatellos:2009a,
  Zhu:2012}.
A disk origin for the triple star system and close-in planets is also
consistent with the fact that the planetary and A--BC orbits seem to
be coplanar, although other primordial formation modes could
potentially bring the protoplanetary disk into alignment with the
triple within the lifetime of the disk \citep{Bate:2000a}.

The key challenge to this disk fragmentation scenario is that
observations indicate that such massive disks are not typical
\citep[e.g.,][]{2015ApJ...802...77M}, although massive disks with
star-to-disk mass ratios close to unity seem to exist
\citep{2012Natur.492...83T}. Such a scenario is at least plausible for
the \obj\ system, which has a total stellar mass of 1.3\,\Msun, as
\citet{2008ApJ...681..375K} showed that protostellar cores with
sufficient mass to produce bound system masses of about 1.5\,\Msun\
are marginally unstable to disk fragmentation.

\section{Conclusions}

We present the first results from our Keck/NIRC2 AO astrometry program
investigating the orbital architectures of planet hosting multiple
systems.  \obj\ is a hierarchical triple star system with five
sub-Earth sized planets in orbit about the primary star.  Combining our
Keck/NIRC2 astrometry with Keck/HIRES RVs of all three stellar
components, we determine that the orbit of the center of mass of
\obj{BC} about \obj{A} is highly eccentric ($e=0.864\pm0.023$) with a
pericenter passage of only $5.0^{+0.9}_{-1.0}$\,AU.  We also find that
this stellar orbit is consistent with being edge-on within the
measurement uncertainties ($i=90.4^{+3.4}_{-3.6}$\,deg), making it
very likely to be coplanar with the planetary system given the low
probability of a misaligned orbit appearing to be this close to
edge-on.  Through direct n-body integrations we validate that this
orbital configuration is dynamically stable both for the triple star
system and five planet system.

We consider a variety of formation scenarios that can simultaneously
explain both the origin of the stellar system (total mass 1.3\,\Msun)
and the existence of the tiny planets on small orbits (total mass
$\approx$1.5\,\Rearth).  We conclude that:
\begin{itemize}

\item The stellar orbit is most likely to be primordial, i.e., in
  place at or before the epoch of planet formation in the system.

\item The protoplanetary disk from which the planets formed would have
  been truncated at 1--2\,AU, severely depleting the reservoir of
  solid material available to form the observed planets.  This
  truncation would have occurred near the ice line, removing most if
  not all volatiles from the descendant planets.

\item The small masses are consistent with some in-situ planet
  formation models, for example the ``inside out'' model.
  This system reinforces the idea that more typical Kepler systems,
  possessing larger planets than seen here, are built by accumulation
  of solids
  drifting in from large disk radii. For the truncated disk of \obj, a
  MMSN scaled up by 20$\times$
  would have adequate solid mass within 2\,AU even at low metallicity
  (80\,\Mearth) to produce these planets at 2\% efficiency.

\item If the natal disk was indeed this massive, the outer regions of
  the disk would have originally been unstable to gravitational
  fragmentation.  Therefore, if such a massive disk is needed to solve
  the mass budget problem, then the triple system might have naturally
  arisen from disk fragments that rapidly evolved dynamically into a
  highly eccentric, coplanar orbital configuration, consistent with
  our observations.

\end{itemize}

The \obj\ system would appear to be a hostile enviroment in which to
form planets.  The host star is metal poor, [Fe/H] = $-0.55$\,dex, and
we have shown that the stellar companions \obj{BC} would have severely
inhibited planet formation in the protoplanetary disk.  Yet five
planets, albeit small ones, did form here.  This may imply that the
assembly of sub-Earth-sized planets is quite robust, particularly in
single star systems with meager disks possessing a reduced supply of
solids.  However, if our suggested formation pathway requiring an
unusually massive disk is correct, then the planet outcome seen here
may not be typical of most multiple star systems that have companions
on solar system scales ($\sim$10--100\,AU).  Future work to build a
large sample of stellar orbit determinations in planet-hosting binary
systems will be key to better understand these fundamental issues for
planet formation.


\acknowledgments

This work was supported by a NASA Keck PI Data Award, administered by
the NASA Exoplanet Science Institute. KMK was supported by NSF
AST-1410174.  We thank Will Best for assistance with some Keck/NIRC2
observations.  It is a pleasure to thank Joel Aycock, Carolyn Jordan,
Jason McIlroy, Luca Rizzi, Terry Stickel, Hien Tran,
and the Keck Observatory staff for assistance with our Keck AO
observations.  
The HIRES data presented here were obtained in collaboration with
Geoff Marcy.
We also thank
James R.\ A.\ Davenport for distributing his IDL implementation
of the cubehelix color scheme \citep{2011BASI...39..289G}.
Our research has employed 
the 2MASS data products; 
NASA's Astrophysical Data System;
and the SIMBAD database operated at CDS, Strasbourg, France.
Finally, the authors wish to recognize and acknowledge the very
significant cultural role and reverence that the summit of Mauna Kea has
always had within the indigenous Hawaiian community.  We are most
fortunate to have the opportunity to conduct observations from this
mountain.

{\it Facilities:} \facility{Keck:II (NGS AO, NIRC2)}

\clearpage

\clearpage

\clearpage

\clearpage

\clearpage

\begin{landscape}
\begin{figure}
  \centerline{\includegraphics[width=4.0in,angle=0]{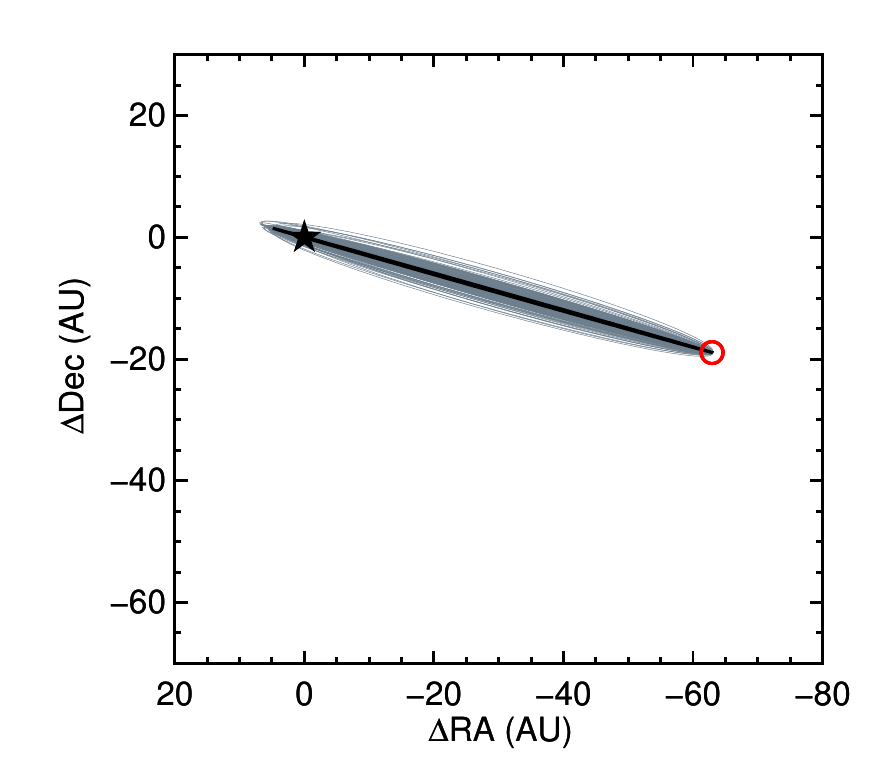}
    \includegraphics[width=4.0in,angle=0]{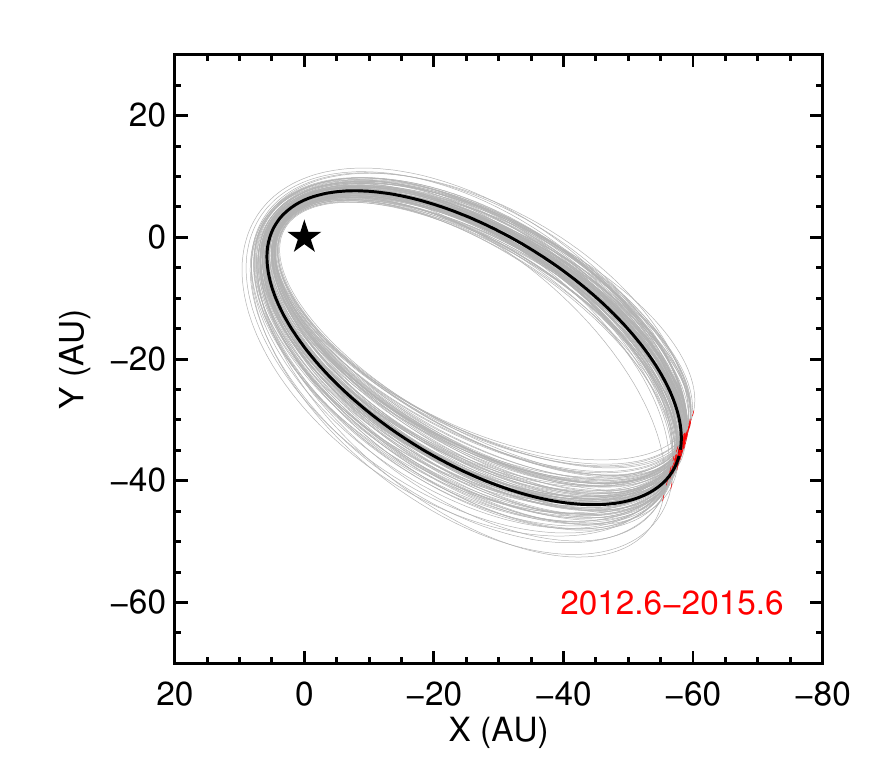}}
  \caption{\normalsize The orbit of the \obj{BC} system in the frame
    of the planet host star \obj{A} (black star).  Our best-fit orbit
    is shown in black, and 100 randomly drawn orbits from our MCMC
    analysis are shown in gray.  Orbit locations that correspond to
    the range of our observation epochs are shown in red.  Left: the
    orbit in plane of the sky, which is consistent with being seen
    edge on. Right: the same orbits shown deprojected in a top down
    view of the orbital plane.  The orbit is currently close to
    apoastron with almost no motion in the plane of the
    sky.  \label{fig:orbit}}
\end{figure}
\end{landscape}

\clearpage

\clearpage

\begin{deluxetable}{lccc}
\tablewidth{0pt}
\tablecaption{Keck/NIRC2 NGS AO Astrometry for \obj{AB} \label{tbl:astrom}}
\tablehead{
\colhead{Date (UT)}                   &
\colhead{Filter}                      &
\colhead{Separation (mas)}            &
\colhead{PA (deg)}                    }
\startdata
2013~Aug~7   & \Kp &  $1842.75\pm0.10$ & $253.266\pm0.028$ \\
2014~Jul~28  & \Kc &  $1842.80\pm0.18$ & $253.279\pm0.008$ \\
2014~Nov~30  & \Kc &  $1841.51\pm0.20$ & $253.284\pm0.005$ \\
2015~Apr~11  & \Kc &  $1841.24\pm0.14$ & $253.247\pm0.007$ \\
2015~Apr~11  & \Kc &  $1841.08\pm0.21$ & $253.274\pm0.013$ \\
\enddata
\tablecomments{Uncertainties quoted here are simply the rms of
  measurements obtained from individual images at each epoch and do
  not account for potential systematic errors (e.g., due to
  uncertainties in the distortion correction, pixel scale and
  orientation).  There are two distinct measurements at the
  2015~Apr~11 epoch because data were obtained at different NIRC2
  orientations placing the binary components on different pixel
  positions on the detector and thus experiencing different distortion
  offsets.}
\end{deluxetable}

\begin{deluxetable}{lr}
\tablewidth{0pt}
\tablecaption{Keck/HIRES Relative Radial Velocities for \obj{A} \label{tbl:rva}}
\tablehead{
\colhead{Date (JD)} &
\colhead{RV$_{\rm A}$ (\ms)}  }
\startdata
2456109.920 & $  8.17\pm1.2$ \\
2456110.831 & $ 10.13\pm1.3$ \\
2456111.897 & $  5.57\pm1.1$ \\
2456112.869 & $  8.11\pm1.3$ \\
2456113.809 & $ 13.53\pm1.2$ \\
2457151.099 & $-12.17\pm1.2$ \\
2457180.106 & $-15.00\pm1.3$ \\
2457180.109 & $-13.76\pm1.2$ \\
2457229.929 & $-13.25\pm1.3$ \\
2457229.932 & $-16.68\pm1.2$ \\
\enddata
\end{deluxetable}

\begin{deluxetable}{lcc}
\tablewidth{0pt}
\tablecaption{Keck/HIRES Radial Velocities for \obj{BC} \label{tbl:rvbc}}
\tablehead{
\colhead{Date (JD)}          &
\colhead{RV$_{\rm B}$ (\kms)}  &
\colhead{RV$_{\rm C}$ (\kms)}  }
\startdata
2456532.7 & $-114.6\pm0.4$ & $-132.8\pm0.4$ \\
2456845.0 & $-133.9\pm0.4$ & $-110.1\pm0.4$ \\
2457229.9 & $-129.5\pm0.4$ & $-116.0\pm0.4$ \\
\enddata
\tablecomments{Uncertainties quoted here were determined from the rms
  of our fit of RV$_{\rm B}$ as a function of RV$_{\rm C}$.}
\end{deluxetable}

\clearpage

\begin{deluxetable}{lcc}
\tablewidth{0pt}
\tablecaption{MCMC Results for the Orbit of \obj{AB} \label{tbl:mcmc}}
\tablehead{
\colhead{Property}              &
\colhead{Median $\pm$1$\sigma$} &
\colhead{Notes/Prior}                  }
\startdata

\multicolumn{3}{c}{Input measurements} \\
\cline{1-3}
Separation, $\rho$ (mas)                                   &  $1843.0\pm0.4$\phn\phn\phn & A \\
$\dot{\rho}$ (\masyr)                                      &    $-1.0\pm0.3$\phs         & A \\
PA, $\theta$ (deg)                                         & $253.258\pm0.021$\phn\phn   & A \\
$\dot{\theta}$ (\degyr)                                    &  $-0.002\pm0.017$\phs       & A \\
$\Delta{\rm RV}_{\rm BC-A}$ (\kms)                        &    $-1.7\pm0.2$\phs         & B \\
$\dot{\rm RV_{\rm A}}$ (\msyr)                             & $-7.8\pm0.5$\phs            & C \\
$\dot{\rm RV_{\rm A}}$ (\msyr)                             & $-11\pm5$\phs\phn           & D \\
System mass, $\Mtot$ (\Msun)                               & $1.30\pm0.06$               & E \\
A--BC mass ratio, $(M_{\rm B}+M_{\rm C})/M_{\rm A}$        &  0.71 (fixed)               & E \\
Distance, $d$ (pc)                                         &  35.7 (fixed)               & \Hipparcos \\

\cline{1-3}
\multicolumn{3}{c}{} \\
\multicolumn{3}{c}{Output posteriors} \\
\cline{1-3}
Semimajor axis, $a$ (AU)                                   & $36.7^{+0.7}_{-0.9}$  & $1/a$ (log-flat) \\
Eccentricity, $e$                                          & $0.864\pm0.023$       & uniform          \\
Inclination, $i$ (deg)                                     & $90.4^{+3.4}_{-3.6}$  & $\sin(i)$        \\
PA of the ascending node, $\Omega$ (deg)                   & $73.1\pm0.9$          & uniform          \\
Argument of periastron, $\omega$ (deg)                     & $342.8^{+3.2}_{-2.6}$ & uniform          \\
Mean longitude at 2456511.83~JD, $\lambda_{\rm ref}$ (deg) & $183.5\pm1.7$         & uniform          \\
Period, $P=\sqrt{a^3/\Mtot}$ (yr)                          & $198^{+8}_{-9}$       & \nodata          \\
Time of periastron, $T_0 = t_{\rm ref} - (\lambda_{\rm ref}-\omega)P/2\pi$ (JD)                             & $2488500\pm900$       & \nodata          \\
Closest approach during periastron, $a(1-e)$ (AU)          & $5.0^{+0.9}_{-1.0}$   & \nodata          \\

\enddata
\tablecomments{The reference epoch for $\lambda_{\rm ref}$ is $t_{\rm
    ref} = 2456511.83$~JD (2013~Aug~7~UT).  Notes on input
  measurements: (A) astrometry corresponds to the epoch range
  2456511.83~JD to 2457124.13~JD; (B) $\Delta{\rm RV}_{\rm BC-A}$
  corresponds to epoch 2456532.74~JD; (C) our HIRES linear RV trend
  corresponds to the epoch range 2456109.92~JD to 2457229.93~JD; (D)
  RV trend from \citet{2009ApJ...697..544S} corresponds to the epoch
  range 2452812.02~JD to 2453568.85~JD; (E)
  \citet{2015ApJ...799..170C} give a mass for \obj{A} of
  $0.76\pm0.04$\,\Msun\ from their asteroseismic analysis, and we
  estimate masses for \obj{B} and C of $0.29\pm0.03$\,\Msun\ and
  $0.25\pm0.03$\,\Msun, respectively, from the
  \citet{2000A&A...364..217D} mass--magnitude relation.}
\end{deluxetable}
\clearpage


\clearpage

\begin{thebibliography}{79}
\expandafter\ifx\csname natexlab\endcsname\relax\def\natexlab#1{#1}\fi

\bibitem[{Adams {et~al.}(2012)Adams, {Ciardi}, {Dupree}, {Gautier}, {Kulesa},
  \& {McCarthy}}]{2012AJ....144...42A}
Adams, E.~R., {Ciardi}, D.~R., {Dupree}, A.~K., {Gautier}, III, T.~N.,
  {Kulesa}, C., \& {McCarthy}, D. 2012, \aj, 144, 42

\bibitem[{Adams {et~al.}(2013)Adams, {Dupree}, {Kulesa}, \&
  {McCarthy}}]{2013AJ....146....9A}
Adams, E.~R., {Dupree}, A.~K., {Kulesa}, C., \& {McCarthy}, D. 2013, \aj, 146,
  9

\bibitem[{{Adams} {et~al.}(2006){Adams}, {Proszkow}, {Fatuzzo}, \&
  {Myers}}]{2006ApJ...641..504A}
{Adams}, F.~C., {Proszkow}, E.~M., {Fatuzzo}, M., \& {Myers}, P.~C. 2006, \apj,
  641, 504

\bibitem[{{Adams} {et~al.}(1989){Adams}, {Ruden}, \&
  {Shu}}]{1989ApJ...347..959A}
{Adams}, F.~C., {Ruden}, S.~P., \& {Shu}, F.~H. 1989, \apj, 347, 959

\bibitem[{{Andrews} {et~al.}(2013){Andrews}, {Rosenfeld}, {Kraus}, \&
  {Wilner}}]{Andrews:2013}
{Andrews}, S.~M., {Rosenfeld}, K.~A., {Kraus}, A.~L., \& {Wilner}, D.~J. 2013,
  \apj, 771, 129

\bibitem[{{Artymowicz} \& {Lubow}(1994)}]{Artymowicz:1994}
{Artymowicz}, P., \& {Lubow}, S.~H. 1994, \apj, 421, 651

\bibitem[{{Bate} {et~al.}(2000){Bate}, {Bonnell}, {Clarke}, {Lubow}, {Ogilvie},
  {Pringle}, \& {Tout}}]{Bate:2000a}
{Bate}, M.~R., {Bonnell}, I.~A., {Clarke}, C.~J., {Lubow}, S.~H., {Ogilvie},
  G.~I., {Pringle}, J.~E., \& {Tout}, C.~A. 2000, \mnras, 317, 773

\bibitem[{{Bechter} {et~al.}(2014){Bechter}, {Crepp}, {Ngo}, {Knutson},
  {Batygin}, {Hinkley}, {Muirhead}, {Johnson}, {Howard}, {Montet}, {Matthews},
  \& {Morton}}]{2014ApJ...788....2B}
{Bechter}, E.~B., {et~al.} 2014, \apj, 788, 2

\bibitem[{{Beck} {et~al.}(2012){Beck}, {Bary}, {Dutrey}, {Pi{\'e}tu},
  {Guilloteau}, {Lubow}, \& {Simon}}]{2012ApJ...754...72B}
{Beck}, T.~L., {Bary}, J.~S., {Dutrey}, A., {Pi{\'e}tu}, V., {Guilloteau}, S.,
  {Lubow}, S.~H., \& {Simon}, M. 2012, \apj, 754, 72

\bibitem[{Borucki {et~al.}(2010)Borucki, {Koch}, {Basri}, {Batalha}, {Brown},
  {Caldwell}, {Caldwell}, {Christensen-Dalsgaard}, {Cochran}, {DeVore},
  {Dunham}, {Dupree}, {Gautier}, {Geary}, {Gilliland}, {Gould}, {Howell},
  {Jenkins}, {Kondo}, {Latham}, {Marcy}, {Meibom}, {Kjeldsen}, {Lissauer},
  {Monet}, {Morrison}, {Sasselov}, {Tarter}, {Boss}, {Brownlee}, {Owen},
  {Buzasi}, {Charbonneau}, {Doyle}, {Fortney}, {Ford}, {Holman}, {Seager},
  {Steffen}, {Welsh}, {Rowe}, {Anderson}, {Buchhave}, {Ciardi}, {Walkowicz},
  {Sherry}, {Horch}, {Isaacson}, {Everett}, {Fischer}, {Torres}, {Johnson},
  {Endl}, {MacQueen}, {Bryson}, {Dotson}, {Haas}, {Kolodziejczak}, {Van Cleve},
  {Chandrasekaran}, {Twicken}, {Quintana}, {Clarke}, {Allen}, {Li}, {Wu},
  {Tenenbaum}, {Verner}, {Bruhweiler}, {Barnes}, \&
  {Prsa}}]{2010Sci...327..977B}
Borucki, W.~J., {et~al.} 2010, Science, 327, 977

\bibitem[{Butler {et~al.}(2006)Butler, {Wright}, {Marcy}, {Fischer}, {Vogt},
  {Tinney}, {Jones}, {Carter}, {Johnson}, {McCarthy}, \&
  {Penny}}]{2006ApJ...646..505B}
Butler, R.~P., {et~al.} 2006, \apj, 646, 505

\bibitem[{Campante {et~al.}(2015)Campante, {Barclay}, {Swift}, {Huber},
  {Adibekyan}, {Cochran}, {Burke}, {Isaacson}, {Quintana}, {Davies}, {Silva
  Aguirre}, {Ragozzine}, {Riddle}, {Baranec}, {Basu}, {Chaplin},
  {Christensen-Dalsgaard}, {Metcalfe}, {Bedding}, {Handberg}, {Stello},
  {Brewer}, {Hekker}, {Karoff}, {Kolbl}, {Law}, {Lundkvist}, {Miglio}, {Rowe},
  {Santos}, {Van Laerhoven}, {Arentoft}, {Elsworth}, {Fischer}, {Kawaler},
  {Kjeldsen}, {Lund}, {Marcy}, {Sousa}, {Sozzetti}, \&
  {White}}]{2015ApJ...799..170C}
Campante, T.~L., {et~al.} 2015, \apj, 799, 170

\bibitem[{{Chambers} {et~al.}(1996){Chambers}, {Wetherill}, \&
  {Boss}}]{Chambers:1996}
{Chambers}, J.~E., {Wetherill}, G.~W., \& {Boss}, A.~P. 1996, \icarus, 119, 261

\bibitem[{{Chatterjee} \& {Tan}(2014)}]{2014ApJ...780...53C}
{Chatterjee}, S., \& {Tan}, J.~C. 2014, \apj, 780, 53

\bibitem[{{Chiang} \& {Laughlin}(2013)}]{2013MNRAS.431.3444C}
{Chiang}, E., \& {Laughlin}, G. 2013, \mnras, 431, 3444

\bibitem[{{Cieza} {et~al.}(2009){Cieza}, {Padgett}, {Allen}, {McCabe},
  {Brooke}, {Carey}, {Chapman}, {Fukagawa}, {Huard}, {Noriga-Crespo},
  {Peterson}, \& {Rebull}}]{2009ApJ...696L..84C}
{Cieza}, L.~A., {et~al.} 2009, \apjl, 696, L84

\bibitem[{Delfosse {et~al.}(2000)Delfosse, {Forveille}, {S{\'e}gransan},
  {Beuzit}, {Udry}, {Perrier}, \& {Mayor}}]{2000A&A...364..217D}
Delfosse, X., {Forveille}, T., {S{\'e}gransan}, D., {Beuzit}, J.-L., {Udry},
  S., {Perrier}, C., \& {Mayor}, M. 2000, \aap, 364, 217

\bibitem[{Diolaiti {et~al.}(2000)Diolaiti, {Bendinelli}, {Bonaccini}, {Close},
  {Currie}, \& {Parmeggiani}}]{2000A&AS..147..335D}
Diolaiti, E., {Bendinelli}, O., {Bonaccini}, D., {Close}, L., {Currie}, D., \&
  {Parmeggiani}, G. 2000, \aaps, 147, 335

\bibitem[{{Doyle} {et~al.}(2011){Doyle}, {Carter}, {Fabrycky}, {Slawson},
  {Howell}, {Winn}, {Orosz}, {Pr{\v s}a}, {Welsh}, {Quinn}, {Latham}, {Torres},
  {Buchhave}, {Marcy}, {Fortney}, {Shporer}, {Ford}, {Lissauer}, {Ragozzine},
  {Rucker}, {Batalha}, {Jenkins}, {Borucki}, {Koch}, {Middour}, {Hall},
  {McCauliff}, {Fanelli}, {Quintana}, {Holman}, {Caldwell}, {Still},
  {Stefanik}, {Brown}, {Esquerdo}, {Tang}, {Furesz}, {Geary}, {Berlind},
  {Calkins}, {Short}, {Steffen}, {Sasselov}, {Dunham}, {Cochran}, {Boss},
  {Haas}, {Buzasi}, \& {Fischer}}]{2011Sci...333.1602D}
{Doyle}, L.~R., {et~al.} 2011, Science, 333, 1602

\bibitem[{Dupuy {et~al.}(2010)Dupuy, {Liu}, {Bowler}, {Cushing}, {Helling},
  {Witte}, \& {Hauschildt}}]{2010ApJ...721.1725D}
Dupuy, T.~J., {Liu}, M.~C., {Bowler}, B.~P., {Cushing}, M.~C., {Helling}, C.,
  {Witte}, S., \& {Hauschildt}, P. 2010, \apj, 721, 1725

\bibitem[{Dupuy {et~al.}(2009)Dupuy, {Liu}, \& {Ireland}}]{2009ApJ...692..729D}
Dupuy, T.~J., {Liu}, M.~C., \& {Ireland}, M.~J. 2009, \apj, 692, 729

\bibitem[{Dupuy {et~al.}(2014)Dupuy, {Liu}, \& {Ireland}}]{2014ApJ...790..133D}
---. 2014, \apj, 790, 133

\bibitem[{{Eastman} {et~al.}(2015){Eastman}, {Beatty}, {Siverd}, {Antognini},
  {Penny}, {Gonzales}, {Crepp}, {Howard}, {Avril}, {Bieryla}, {Collins},
  {Fulton}, {Ge}, {Gregorio}, {Ma}, {Mellon}, {Oberst}, {Wang}, {Gaudi},
  {Pepper}, {Stassun}, {Buchhave}, {Jensen}, {Latham}, {Berlind}, {Calkins},
  {Cargile}, {Colon}, {Dhital}, {Esquerdo}, {Johnson}, {Kielkopf}, {Manner},
  {Mao}, {McLeod}, {Penev}, {Stefanik}, {Street}, {Zambelli}, {DePoy}, {Gould},
  {Marshall}, {Pogge}, {Trueblood}, \& {Trueblood}}]{2015arXiv151000015E}
{Eastman}, J.~D., {et~al.} 2015, ArXiv e-prints

\bibitem[{Everett {et~al.}(2015)Everett, {Barclay}, {Ciardi}, {Horch},
  {Howell}, {Crepp}, \& {Silva}}]{2015AJ....149...55E}
Everett, M.~E., {Barclay}, T., {Ciardi}, D.~R., {Horch}, E.~P., {Howell},
  S.~B., {Crepp}, J.~R., \& {Silva}, D.~R. 2015, \aj, 149, 55

\bibitem[{{Everhart}(1985)}]{1985dcto.proc..185E}
{Everhart}, E. 1985, in Dynamics of Comets: Their Origin and Evolution,
  Proceedings of IAU Colloq. 83, held in Rome, Italy, June 11-15, 1984. Edited
  by Andrea Carusi and Giovanni B. Valsecchi. Dordrecht: Reidel, Astrophysics
  and Space Science Library. Volume 115, 1985, p.185, ed. A.~{Carusi} \& G.~B.
  {Valsecchi}, 185

\bibitem[{Foreman-Mackey {et~al.}(2013)Foreman-Mackey, {Hogg}, {Lang}, \&
  {Goodman}}]{2013PASP..125..306F}
Foreman-Mackey, D., {Hogg}, D.~W., {Lang}, D., \& {Goodman}, J. 2013, \pasp,
  125, 306

\bibitem[{{Ghez} {et~al.}(1997){Ghez}, {White}, \&
  {Simon}}]{1997ApJ...490..353G}
{Ghez}, A.~M., {White}, R.~J., \& {Simon}, M. 1997, \apj, 490, 353

\bibitem[{{Green}(2011)}]{2011BASI...39..289G}
{Green}, D.~A. 2011, Bulletin of the Astronomical Society of India, 39, 289

\bibitem[{{Hamers} \& {Portegies Zwart}(2015)}]{2015arXiv151100944H}
{Hamers}, A.~S., \& {Portegies Zwart}, S.~F. 2015, ArXiv e-prints

\bibitem[{{Hansen} \& {Murray}(2012)}]{2012ApJ...751..158H}
{Hansen}, B.~M.~S., \& {Murray}, N. 2012, \apj, 751, 158

\bibitem[{{Harris} {et~al.}(2012){Harris}, {Andrews}, {Wilner}, \&
  {Kraus}}]{2012ApJ...751..115H}
{Harris}, R.~J., {Andrews}, S.~M., {Wilner}, D.~J., \& {Kraus}, A.~L. 2012,
  \apj, 751, 115

\bibitem[{{Hatzes} {et~al.}(2003){Hatzes}, {Cochran}, {Endl}, {McArthur},
  {Paulson}, {Walker}, {Campbell}, \& {Yang}}]{2003ApJ...599.1383H}
{Hatzes}, A.~P., {Cochran}, W.~D., {Endl}, M., {McArthur}, B., {Paulson},
  D.~B., {Walker}, G.~A.~H., {Campbell}, B., \& {Yang}, S. 2003, \apj, 599,
  1383

\bibitem[{{Hayashi}(1981)}]{1981PThPS..70...35H}
{Hayashi}, C. 1981, Progress of Theoretical Physics Supplement, 70, 35

\bibitem[{{Heppenheimer}(1974)}]{1974Icar...22..436H}
{Heppenheimer}, T.~A. 1974, \icarus, 22, 436

\bibitem[{{Holman} \& {Wiegert}(1999)}]{1999AJ....117..621H}
{Holman}, M.~J., \& {Wiegert}, P.~A. 1999, \aj, 117, 621

\bibitem[{{Howard} {et~al.}(2010){Howard}, {Johnson}, {Marcy}, {Fischer},
  {Wright}, {Bernat}, {Henry}, {Peek}, {Isaacson}, {Apps}, {Endl}, {Cochran},
  {Valenti}, {Anderson}, \& {Piskunov}}]{2010ApJ...721.1467H}
{Howard}, A.~W., {et~al.} 2010, \apj, 721, 1467

\bibitem[{{Jang-Condell} {et~al.}(2008){Jang-Condell}, {Mugrauer}, \&
  {Schmidt}}]{2008ApJ...683L.191J}
{Jang-Condell}, H., {Mugrauer}, M., \& {Schmidt}, T. 2008, \apjl, 683, L191

\bibitem[{{Kozai}(1962)}]{1962AJ.....67..591K}
{Kozai}, Y. 1962, \aj, 67, 591

\bibitem[{{Kratter} {et~al.}(2008){Kratter}, {Matzner}, \&
  {Krumholz}}]{2008ApJ...681..375K}
{Kratter}, K.~M., {Matzner}, C.~D., \& {Krumholz}, M.~R. 2008, \apj, 681, 375

\bibitem[{{Kratter} {et~al.}(2010{\natexlab{a}}){Kratter}, {Matzner},
  {Krumholz}, \& {Klein}}]{2010ApJ...708.1585K}
{Kratter}, K.~M., {Matzner}, C.~D., {Krumholz}, M.~R., \& {Klein}, R.~I.
  2010{\natexlab{a}}, \apj, 708, 1585

\bibitem[{{Kratter} {et~al.}(2010{\natexlab{b}}){Kratter}, {Murray-Clay}, \&
  {Youdin}}]{2010ApJ...710.1375K}
{Kratter}, K.~M., {Murray-Clay}, R.~A., \& {Youdin}, A.~N. 2010{\natexlab{b}},
  \apj, 710, 1375

\bibitem[{{Kraus} {et~al.}(2012){Kraus}, {Ireland}, {Hillenbrand}, \&
  {Martinache}}]{2012ApJ...745...19K}
{Kraus}, A.~L., {Ireland}, M.~J., {Hillenbrand}, L.~A., \& {Martinache}, F.
  2012, \apj, 745, 19

\bibitem[{{Kraus} {et~al.}(2011){Kraus}, {Ireland}, {Martinache}, \&
  {Hillenbrand}}]{2011ApJ...731....8K}
{Kraus}, A.~L., {Ireland}, M.~J., {Martinache}, F., \& {Hillenbrand}, L.~A.
  2011, \apj, 731, 8

\bibitem[{{Krumholz} {et~al.}(2007){Krumholz}, {Klein}, \&
  {McKee}}]{2007ApJ...656..959K}
{Krumholz}, M.~R., {Klein}, R.~I., \& {McKee}, C.~F. 2007, \apj, 656, 959

\bibitem[{{Lee} {et~al.}(2014){Lee}, {Chiang}, \&
  {Ormel}}]{2014ApJ...797...95L}
{Lee}, E.~J., {Chiang}, E., \& {Ormel}, C.~W. 2014, \apj, 797, 95

\bibitem[{{Levison} \& {Duncan}(2013)}]{2013ascl.soft03001L}
{Levison}, H.~F., \& {Duncan}, M.~J. 2013, {SWIFT: A solar system integration
  software package}, Astrophysics Source Code Library

\bibitem[{{Li} {et~al.}(2014){Li}, {Naoz}, {Kocsis}, \&
  {Loeb}}]{2014ApJ...785..116L}
{Li}, G., {Naoz}, S., {Kocsis}, B., \& {Loeb}, A. 2014, \apj, 785, 116

\bibitem[{{Lidov}(1962)}]{1962P&SS....9..719L}
{Lidov}, M.~L. 1962, \planss, 9, 719

\bibitem[{Lillo-Box {et~al.}(2014)Lillo-Box, {Barrado}, \&
  {Bouy}}]{2014A&A...566A.103L}
Lillo-Box, J., {Barrado}, D., \& {Bouy}, H. 2014, \aap, 566, A103

\bibitem[{{Lissauer}(1987)}]{Lissauer:1987}
{Lissauer}, J.~J. 1987, Icarus, 69, 249

\bibitem[{{Lissauer} {et~al.}(2011){Lissauer}, {Ragozzine}, {Fabrycky},
  {Steffen}, {Ford}, {Jenkins}, {Shporer}, {Holman}, {Rowe}, {Quintana},
  {Batalha}, {Borucki}, {Bryson}, {Caldwell}, {Carter}, {Ciardi}, {Dunham},
  {Fortney}, {Gautier}, {Howell}, {Koch}, {Latham}, {Marcy}, {Morehead}, \&
  {Sasselov}}]{2011ApJS..197....8L}
{Lissauer}, J.~J., {et~al.} 2011, \apjs, 197, 8

\bibitem[{Lucy(2014)}]{2014A&A...563A.126L}
Lucy, L.~B. 2014, \aap, 563, A126

\bibitem[{{Mann} {et~al.}(2015){Mann}, {Andrews}, {Eisner}, {Williams},
  {Meyer}, {Di Francesco}, {Carpenter}, \& {Johnstone}}]{2015ApJ...802...77M}
{Mann}, R.~K., {Andrews}, S.~M., {Eisner}, J.~A., {Williams}, J.~P., {Meyer},
  M.~R., {Di Francesco}, J., {Carpenter}, J.~M., \& {Johnstone}, D. 2015, \apj,
  802, 77

\bibitem[{{Martin} {et~al.}(2012){Martin}, {Lubow}, {Livio}, \&
  {Pringle}}]{Martin:2012}
{Martin}, R.~G., {Lubow}, S.~H., {Livio}, M., \& {Pringle}, J.~E. 2012, \mnras,
  420, 3139

\bibitem[{{Mohanty} {et~al.}(2013){Mohanty}, {Ercolano}, \&
  {Turner}}]{2013ApJ...764...65M}
{Mohanty}, S., {Ercolano}, B., \& {Turner}, N.~J. 2013, \apj, 764, 65

\bibitem[{{Najita} \& {Kenyon}(2014)}]{2014MNRAS.445.3315N}
{Najita}, J.~R., \& {Kenyon}, S.~J. 2014, \mnras, 445, 3315

\bibitem[{{Perez-Becker} \& {Chiang}(2013)}]{Perez-Becker:2013}
{Perez-Becker}, D., \& {Chiang}, E. 2013, \mnras, 433, 2294

\bibitem[{{Pineda} {et~al.}(2015){Pineda}, {Offner}, {Parker}, {Arce},
  {Goodman}, {Caselli}, {Fuller}, {Bourke}, \& {Corder}}]{2015Natur.518..213P}
{Pineda}, J.~E., {et~al.} 2015, \nat, 518, 213

\bibitem[{{Rafikov}(2015)}]{Rafikov:2015}
{Rafikov}, R.~R. 2015, \apj, 804, 62

\bibitem[{{Schlichting}(2014)}]{2014ApJ...795L..15S}
{Schlichting}, H.~E. 2014, \apjl, 795, L15

\bibitem[{{Silsbee} \& {Rafikov}(2015)}]{2015ApJ...798...71S}
{Silsbee}, K., \& {Rafikov}, R.~R. 2015, \apj, 798, 71

\bibitem[{Sozzetti {et~al.}(2009)Sozzetti, {Torres}, {Latham}, {Stefanik},
  {Korzennik}, {Boss}, {Carney}, \& {Laird}}]{2009ApJ...697..544S}
Sozzetti, A., {Torres}, G., {Latham}, D.~W., {Stefanik}, R.~P., {Korzennik},
  S.~G., {Boss}, A.~P., {Carney}, B.~W., \& {Laird}, J.~B. 2009, \apj, 697, 544

\bibitem[{{Stamatellos} \& {Whitworth}(2009)}]{Stamatellos:2009a}
{Stamatellos}, D., \& {Whitworth}, A.~P. 2009, \mnras, 392, 413

\bibitem[{{Stone} {et~al.}(2014){Stone}, {Eisner}, {Salyk}, {Kulesa}, \&
  {McCarthy}}]{2014ApJ...792...56S}
{Stone}, J.~M., {Eisner}, J.~A., {Salyk}, C., {Kulesa}, C., \& {McCarthy}, D.
  2014, \apj, 792, 56

\bibitem[{{Tobin} {et~al.}(2012){Tobin}, {Hartmann}, {Chiang}, {Wilner},
  {Looney}, {Loinard}, {Calvet}, \& {D'Alessio}}]{2012Natur.492...83T}
{Tobin}, J.~J., {Hartmann}, L., {Chiang}, H.-F., {Wilner}, D.~J., {Looney},
  L.~W., {Loinard}, L., {Calvet}, N., \& {D'Alessio}, P. 2012, \nat, 492, 83

\bibitem[{{Valtonen} \& {Karttunen}(2006)}]{2006tbp..book.....V}
{Valtonen}, M., \& {Karttunen}, H. 2006, {The Three-Body Problem}

\bibitem[{{Valtonen} {et~al.}(2008){Valtonen}, {Myll{\"a}ri}, {Orlov}, \&
  {Rubinov}}]{2008IAUS..246..209V}
{Valtonen}, M., {Myll{\"a}ri}, A., {Orlov}, V., \& {Rubinov}, A. 2008, in IAU
  Symposium, Vol. 246, IAU Symposium, ed. E.~{Vesperini}, M.~{Giersz}, \&
  A.~{Sills}, 209--217

\bibitem[{van Leeuwen(2007)}]{2007hnrr.book.....V}
van Leeuwen, F. 2007, {Hipparcos, the New Reduction of the Raw Data}
  (Hipparcos, the New Reduction of the Raw Data.~By Floor van Leeuwen,
  Institute of Astronomy, Cambridge University, Cambridge, UK Series:
  Astrophysics and Space Science Library, Vol.~ 350 20 Springer Dordrecht)

\bibitem[{Wang {et~al.}(2014)Wang, {Fischer}, {Xie}, \&
  {Ciardi}}]{2014ApJ...791..111W}
Wang, J., {Fischer}, D.~A., {Xie}, J.-W., \& {Ciardi}, D.~R. 2014, \apj, 791,
  111

\bibitem[{Wang {et~al.}(2015)Wang, {Fischer}, {Xie}, \&
  {Ciardi}}]{2015arXiv151001964W}
---. 2015, ArXiv e-prints

\bibitem[{{Weidenschilling}(1977)}]{1977Ap&SS..51..153W}
{Weidenschilling}, S.~J. 1977, \apss, 51, 153

\bibitem[{{Welsh} {et~al.}(2012){Welsh}, {Orosz}, {Carter}, {Fabrycky}, {Ford},
  {Lissauer}, {Pr{\v s}a}, {Quinn}, {Ragozzine}, {Short}, {Torres}, {Winn},
  {Doyle}, {Barclay}, {Batalha}, {Bloemen}, {Brugamyer}, {Buchhave},
  {Caldwell}, {Caldwell}, {Christiansen}, {Ciardi}, {Cochran}, {Endl},
  {Fortney}, {Gautier}, {Gilliland}, {Haas}, {Hall}, {Holman}, {Howard},
  {Howell}, {Isaacson}, {Jenkins}, {Klaus}, {Latham}, {Li}, {Marcy}, {Mazeh},
  {Quintana}, {Robertson}, {Shporer}, {Steffen}, {Windmiller}, {Koch}, \&
  {Borucki}}]{2012Natur.481..475W}
{Welsh}, W.~F., {et~al.} 2012, \nat, 481, 475

\bibitem[{{White} \& {Ghez}(2001)}]{2001ApJ...556..265W}
{White}, R.~J., \& {Ghez}, A.~M. 2001, \apj, 556, 265

\bibitem[{Wilson(1941)}]{1941ApJ....93...29W}
Wilson, O.~C. 1941, \apj, 93, 29

\bibitem[{Wizinowich {et~al.}(2000)}]{2000PASP..112..315W}
Wizinowich, P., {et~al.} 2000, \pasp, 112, 315

\bibitem[{Yelda {et~al.}(2010)Yelda, {Lu}, {Ghez}, {Clarkson}, {Anderson},
  {Do}, \& {Matthews}}]{2010ApJ...725..331Y}
Yelda, S., {Lu}, J.~R., {Ghez}, A.~M., {Clarkson}, W., {Anderson}, J., {Do},
  T., \& {Matthews}, K. 2010, \apj, 725, 331

\bibitem[{{Youdin} \& {Goodman}(2005)}]{Youdin:2005}
{Youdin}, A.~N., \& {Goodman}, J. 2005, \apj, 620, 459

\bibitem[{{Young} \& {Clarke}(2015)}]{2015MNRAS.452.3085Y}
{Young}, M.~D., \& {Clarke}, C.~J. 2015, \mnras, 452, 3085

\bibitem[{{Zhu} {et~al.}(2012){Zhu}, {Hartmann}, {Nelson}, \&
  {Gammie}}]{Zhu:2012}
{Zhu}, Z., {Hartmann}, L., {Nelson}, R.~P., \& {Gammie}, C.~F. 2012, \apj, 746,
  110

\end{thebibliography}
\end{document}